\documentclass[twocolumn]{aastex631}
\usepackage{booktabs}
\usepackage{IEEEtrantools}
\usepackage{graphicx}
\usepackage{enumitem}
\usepackage{amsmath}
\usepackage{comment}
\usepackage{tabularx}
\usepackage{multirow}

\newcommand{\msun}{\,{\rm M}_\odot}
\newcommand{\pc}{\,{\rm pc}}

\newcommand{\Mpc}{\,{\rm Mpc}}
\newcommand{\Gpc}{\,{\rm Gpc}}
\newcommand{\au}{\,{\rm au}}
\newcommand{\km}{\,{\rm km}}
\newcommand{\s}{\,{\rm s}}
\newcommand{\g}{\,{\rm g}}
\newcommand{\cm}{\,{\rm cm}}
\newcommand{\yr}{\,{\rm yr}}
\newcommand{\Myr}{\,{\rm Myr}}
\newcommand{\Gyr}{\,{\rm Gyr}}

\definecolor{sjcol}{rgb}{0.0, 0.5, 0.0}
\definecolor{sjcolrm}{rgb}{1.0, 0.13, 0.32}
\definecolor{ewcol}{rgb}{0.8,0,0}

\usepackage{natbib}
\usepackage{verbatim}
\usepackage{mathptmx}  % for better fonts

\begin{document}

\title{Formation and Evolution of Compact Binaries Containing Intermediate Mass Black Holes in Dense Star Clusters}

\author[0000-0003-4769-6431]{Seungjae Lee}
\correspondingauthor{Seungjae Lee, \href{mailto:balgun1004@gmail.com}{balgun1004@gmail.com}}
\affiliation{Center for Theoretical Physics, Department of Physics and Astronomy, Seoul National University, Seoul 08826, Korea}

\author[0000-0003-4412-7161]{Hyung Mok Lee}
\affiliation{Research Institute for Basic Sciences, Seoul National University, Seoul 08826, Korea}
\affiliation{Seoul National University Astronomy Research Center, Seoul 08826, Korea}

\author[0000-0003-4464-1160]{Ji-hoon Kim}
\correspondingauthor{Ji-hoon Kim, \href{mailto:mornkr@snu.ac.kr}{mornkr@snu.ac.kr}}
\affiliation{Center for Theoretical Physics, Department of Physics and Astronomy, Seoul National University, Seoul 08826, Korea}
\affiliation{Seoul National University Astronomy Research Center, Seoul 08826, Korea}
\affiliation{Institute for Data Innovation in Science, Soul National University, Seoul 08826, Korea}

\author[0000-0003-2264-7203]{Rainer Spurzem}
\affiliation{Astronomisches Rechen-Institut, Zentrum f\"ur Astronomie der Universit\"at Heidelberg, M\"onchhofstr. 12-14, D-69120 Heidelberg, Germany}
\affiliation{National Astronomical Observatories and Key Laboratory of Computational Astrophysics, Chinese Academy of Sciences, 20A Datun Rd., Chaoyang District, 
Beijing 100101, China}
\affiliation{Kavli Institute for Astronomy and Astrophysics, Peking University, Yiheyuan Lu 5, Haidian Qu, 100871, Beijing, China}

\author[0000-0002-5097-8707]{Jongsuk Hong}
\affiliation{Korea Astronomy and Space Science Institute, Daejeon 34055, Republic of Korea}

\author[0009-0002-3230-8205]{Eunwoo Chung}
\affiliation{Center for Theoretical Physics, Department of Physics and Astronomy, Seoul National University, Seoul 08826, Korea}
% \collaboration{20}{(AAS Journals Data Editors)}



H
\begin{abstract}
We investigate the evolution of star clusters containing intermediate-mass black hole (IMBH) of $300$ to $5000\msun$, focusing on the formation and evolution of IMBH-stellar mass black holes (SBHs; $M_{\rm BH} \lesssim 10^2 \msun$) binaries.
Dense stellar systems like globular clusters (GCs) or nuclear star clusters offer unique laboratories for studying the existence and impact of IMBHs.
IMBHs residing in GCs have been under speculation for decades, with their broad astrophysical implications for the cluster's dynamical evolution, stellar population, GW signatures, among others.
While existing GW observatories such as the Advanced Laser Interferometer Gravitational-wave Observatory (aLIGO) target binaries with relatively modest mass ratios, $q \lesssim 10$, future observatories such as the Einstein Telescope (ET) and the Laser Interferometer Space Antenna (LISA) will detect intermediate-mass ratio inspirals (IMRIs) with $q > 10$.
This work explores the potential for detecting IMRIs adopting these upcoming telescopes.
For our experiments, we perform multiple direct $N$-body simulations with IMBHs utilizing {\tt Nbody6++GPU}, after implementing the GW merger schemes for IMBHs.
We then study the statistical properties of the resulting IMRIs, such as the event rates and orbital properties.
Assuming that IMRIs with a signal-to-noise ratio $S/N > 8$ are detectable, we derive the following detection rates for each observatory: $\lesssim  0.02\yr^{-1}$ for aLIGO, $\sim 101 - 355 \yr^{-1}$ for ET, $\sim 186 - 200 \yr^{-1}$ for LISA, $\sim 0.24 - 0.34 \yr^{-1}$ for aSOGRO, and $\sim 3880 - 4890 \yr^{-1}$ for DECIGO.
Our result confirms the capability of detecting IMRIs with future GW telescopes.
\end{abstract}


\keywords{methods: numerical, galaxies: star clusters, galaxies: supermassive black holes, star clusters: general, gravitational waves, black hole physics}


\section{Introduction} \label{sec:intro} 

Commonly classified as intermediate-mass black holes (IMBHs), black holes (BHs) with masses ranging from $10^2$ to $10^5\msun$, bridge the gap between stellar-mass black holes (SBHs; $M_{\rm BH} \lesssim 10^2\msun$) and the massive black holes (MBHs; $M_{\rm BH} \gtrsim 10^6\msun$), typically found at the galactic centers \citep[for reviews, see e.g.,][]{Greene2020ARA&A..58..257G}.
IMBHs may be the key to understanding the origin of MBHs and play critical roles in various astrophysical environments.
However, their definitive detection and characterization remain exceptionally challenging due to their relatively modest gravitational influence and the limitations of current observational resolution.

% IMBH가 Dense star cluster에 있을 가능성이 높은 이유.
Dense star clusters such as globular clusters (GCs; $M_{\rm GC} \sim 10^3 - 10^6\msun$) and nuclear star clusters (NSCs; $M_{\rm NSC} \sim 10^6 - 10^8\msun$) have long been considered as the possible habitats for IMBHs.
NSCs, the most massive stellar system at galactic centers \citep[for reviews, see e.g.,][]{Neumayer2020A&ARv..28....4N}, are known for their co-existence with MBHs \citep[e.g.,][]{Bacon1994A&A...281..691B, Nguyen2018ApJ...858..118N}.
Some studies such as \citet{Antonini2015ApJ...812...72A} have proposed that MBHs might be correlated with their host NSCs.
A naive generalization of this result leads us to speculate that the IMBHs can be populated in GCs or lower-mass NSCs.

If IMBHs do exist within GCs, there might be multiple possible pathways toward their coexistence.
One possibility is that the IMBHs form via frequent dynamical interactions between stars and BHs within dense star clusters like GCs.
If the core of a star cluster without a central massive object quickly contracts because of the gravothermal catastrophe \citep{Cohn1980ApJ...242..765C, Takahashi1995PASJ...47..561T}, it could lead to the formation of a very massive star (VMS) through frequent star-star collisions, eventually collapsing into a BH with a mass above the upper mass gap \citep{Begelman1978MNRAS.185..847B, Portegies2002ApJ...576..899P, Gurkan2004ApJ...604..632G, Freitag2006MNRAS.368..121F, Freitag2006MNRAS.368..141F, Giersz2015MNRAS.454.3150G, Mapelli2016MNRAS.459.3432M}. 
Alternatively, a stellar-mass BH might grow through multiple interactions with stars and other stellar-mass BHs \citep{Giersz2015MNRAS.454.3150G, Rizzuto2021MNRAS.501.5257R, Gonzalez2021ApJ...908L..29G}.
Another possibility is that a wandering IMBH is captured by a star cluster and eventually settles in its core.
%Regardless of the scenario, many observers have tried to establish the existence of IMBHs.

% IMBH의 간접적인 관측 증거들
Although it is very challenging to detect IMBHs with current observation and measurement techniques, there are several IMBH candidates discovered so far \citep{Greene2020ARA&A..58..257G}.
One example is a luminous X-ray outburst suggested to be a tidal disruption (TD) event induced by IMBH \citep{Lin2018NatAs...2..656L}.
Some studies suggest that there can be IMBHs within GCs such as $\omega$ Cen \citep{Noyola2010ApJ...719L..60N, Baumgardt2017MNRAS.464.2174B}.
It is recently backed up by the fast-moving stars in the innermost area ($\lesssim 0.08\pc$) of the star cluster \citep{Haberle2024Natur.631..285H}.
Additionally, IMBHs can be populated inside the galactic nuclei of low-mass galaxies \citep{dan2015ApJ...809..101D, Nguyen2019ApJ...872..104N, Woo2019NatAs...3..755W} as an extension of galaxy-MBH relation \citep{Kormendy2013ARA&A..51..511K}.
However, none of these studies have definitively confirmed the existence of IMBHs.

Conclusive evidence for IMBHs with masses $\gtrsim 10^3\msun$ is still lacking in the gravitational wave (GW) as well.
For example, while the GW research has made significant progress in the last few years, most GW signals detected so far have been associated with mergers between BHs with masses $\lesssim 100\msun$.
Currently, the main observational targets of existing GW observatories such as Advanced Laser Interferometer Gravitational-wave Observatory (aLIGO) and Virgo, or the upcoming third-generation Einstein Telescope (ET) are binaries with mass ratios $q \lesssim 10$.
However, they may be capable of detecting some binaries involving IMBHs --- including the intermediate mass ratio inspirals (IMRIs) of mass ratios $q > 10$ \citep{Hild2011CQGra..28i4013H, Sathyaprakash2012CQGra..29l4013S, Amaro2018PhRvD..98f3018A}.
Moreover, the space-based Laser Interferometer Space Antenna (LISA), Decihertz Interferometer Gravitational Wave Observatory (DECIGO), or the Advanced Superconducting Omnidirectional Gravitational Radiation Observatory \citep[aSOGRO;][]{Paik2016CQGra..33g5003P, Bae2024PTEP.2024e3E01B} will be specialized at detecting lower frequency bands.
In addition, those observatories are likely to complement each other due to their distinct detection bands.
These GW signatures may prove to be a promising way to establish the existence of IMBHs.

Numerous works made estimates of the IMRI or their detection rates for decades using semi-analytic or numerical framework.
The early semi-analytic works prospected that the aLIGO would detect IMRIs including IMBHs with masses $M_{\rm IMBH} < 350\msun$ in the rate $\lesssim 10 - 30 \yr^{-1}$ \citep{Brown2007PhRvL..99t1102B, Mandel2008ApJ...681.1431M}.
With a broader IMBH mass range ($\sim 100 - 1000\msun$), the analytic work of \citet{Gair2011GReGr..43..485G} provided the IMRI event rate of $2 - 175 \yr^{-1}$.
Recently, \citet{Fragione2022ApJ...933..170F} proposed that the detection rate of ET would be $0.01 - 10 \Gpc^{-3} \yr^{-1}$ for IMBH with masses $\lesssim 1000\msun$.
In the side of numerical experiments, \citet{Hong2015MNRAS.448..754H} explored the evolution of NSC in the presence of MBHs with GPU accelerated \textsc{\footnotesize NBODY}{6} code \citep{Nitadori2012MNRAS.424..545N}, but they mainly focused on the formation of SBH binaries.
\citet{Arca-Sedda2021A&A...652A..54A} simulated the evolution of three-body systems (one IMBH and two SBHs) within the GC potential.
They obtained the IMRI detection rate ranging from $0.003 \yr^{-1}$ to $3000\yr^{-1}$ which depends on the nature of detectors.
\citet{Wang2022MNRAS.515.5106W} provided the IMRI rates $\sim 0.1 - 0.8 \Gpc^{-3} \yr^{-1}$ for IMBH with masses $M_{\rm IMBH} \lesssim 1000\msun$ in the population III (Pop III) star clusters utilizing a direct {\it N}-body simulation code {\tt PETAR} \citep{Wang2020MNRAS.497..536W}.

In this work, we explore the possibility of detecting IMRIs by the existing and upcoming GW observatories by estimating the detection rate of IMRIs in GCs under the assumption of their coexistence.
We perform a suite of direct {\it N}-body simulations using the code {\tt Nbody6++GPU} \citep{Kamlah2022MNRAS.511.4060K, Spurzem2023LRCA....9....3S} to study the dynamical interactions of an IMBH embedded at the center of a dense stellar system.
Our simulation incorporates the GW merger and TD of the IMBHs, allowing the IMBHs to accrete SBHs and stars nearby.
Also, we examine the IMBH mass range ($300 - 5000 \msun$) broader than most of the precedent studies as LISA is expected to have advantages in detecting IMRIs including an IMBH with mass $\gtrsim 1000\msun$.
Lastly, we investigate the statistics of the resulting GW merger events such as their orbital properties and their detectability by 5 different instruments (aLIGO, ET, LISA, aSOGRO, and DECIGO).

This paper is organized as follows. In Section 2, we briefly describe our code {\tt Nbody6++GPU} and the IMBH model we have implemented, such as GW merger and TD.
Then we provide the initial conditions tested in our studies in Section 3.
We show the result of our simulations including the IMRI rates in Section 4.
Finally, we discuss our result in Section 5 and conclude our study in Section 6.

%\vspace{1mm}

\section{Methods}\label{sec: Method}

\begin{figure*}
    \centering    
    \includegraphics[width=0.85\textwidth]{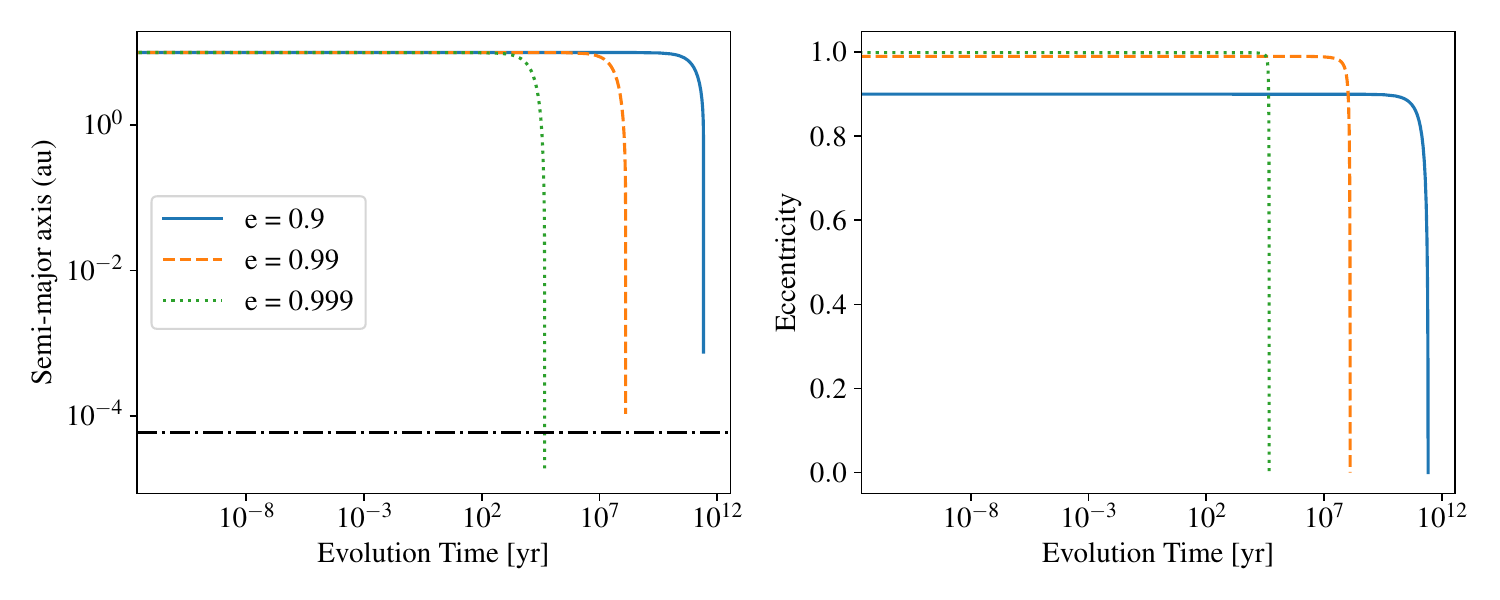}
    \vspace{-2mm}    
    \caption{Comparison of the IMBH-BH binary evolution for different initial eccentricities based on Eqs.(\ref{eq: dadt}) and (\ref{eq: dedt}). 
    All binaries have $M_{\rm IMBH} = 10^3\msun$ and $M_{\rm BH} = 40\msun$, and begin with $a = 10 \au$. 
    {\it Left:} The closer the initial eccentricity of the binary is to 1, the shorter its merger timescale becomes (see Section \ref{sec: snr}). 
    The {\it black dash-dotted line} refers to the Schwarzschild radius of the IMBH. 
    The merger timescales estimated by Eq.(\ref{eq: T_gw}) are $t_{\rm GW} = 2.31\times 10^{11}\yr$, $8.58\times 10^{6}\yr$, and $2.76\times 10^{4}\yr$, for $e=0.9$, 0.99, and 0.999, respectively.
    {\it Right:} The evolution of eccentricity as the binary evolves in time.
    This figure shows the significant impacts of the initial eccentricity significantly on the merger timescale.}
    \label{fig: Evolution of IMRI}
    \vspace{4mm}
\end{figure*}

In this work, we carry out a series of simulations by exploiting the state-of-the-art direct summation $N$-body simulation code with GPU acceleration, {\tt Nbody6++GPU}.
It utilizes a 4th-order Hermite integrator with individual block-time steps and advanced methods for handling close encounters and few-body dynamics \citep{McMillan1986, Hut1995ApJ...443L..93H}. 
These methods include the Kustaanheimo-Stiefel (KS) regularization \citep{KustaanheimoSCHINZELDAVENPORTSTIEFEL+1965+204+219}, the Ahmad-Cohen scheme for neighboring stars \citep{Ahmad1973JCoPh..12..389A}, and algorithmic chain regularization \citep{Mikkola1999MNRAS.310..745M,Mikkola2008AJ....135.2398M}. 
This combination is capable of tracking the detailed evolution of binaries with periods up to $10^{-10}$ times shorter than the dynamical timescales of star clusters.
For details of the methods adopted in {\tt Nbody6++GPU}, we refer the interested readers to \cite{Aarseth1999PASP..111.1333A, Aarseth2003gnbs.book.....A}, \cite{Wang2015MNRAS.450.4070W}, \cite{Kamlah2022MNRAS.511.4060K}, and \cite{Spurzem2023LRCA....9....3S}.

\begin{figure}
    \centering
    \includegraphics[width=1.03\linewidth]{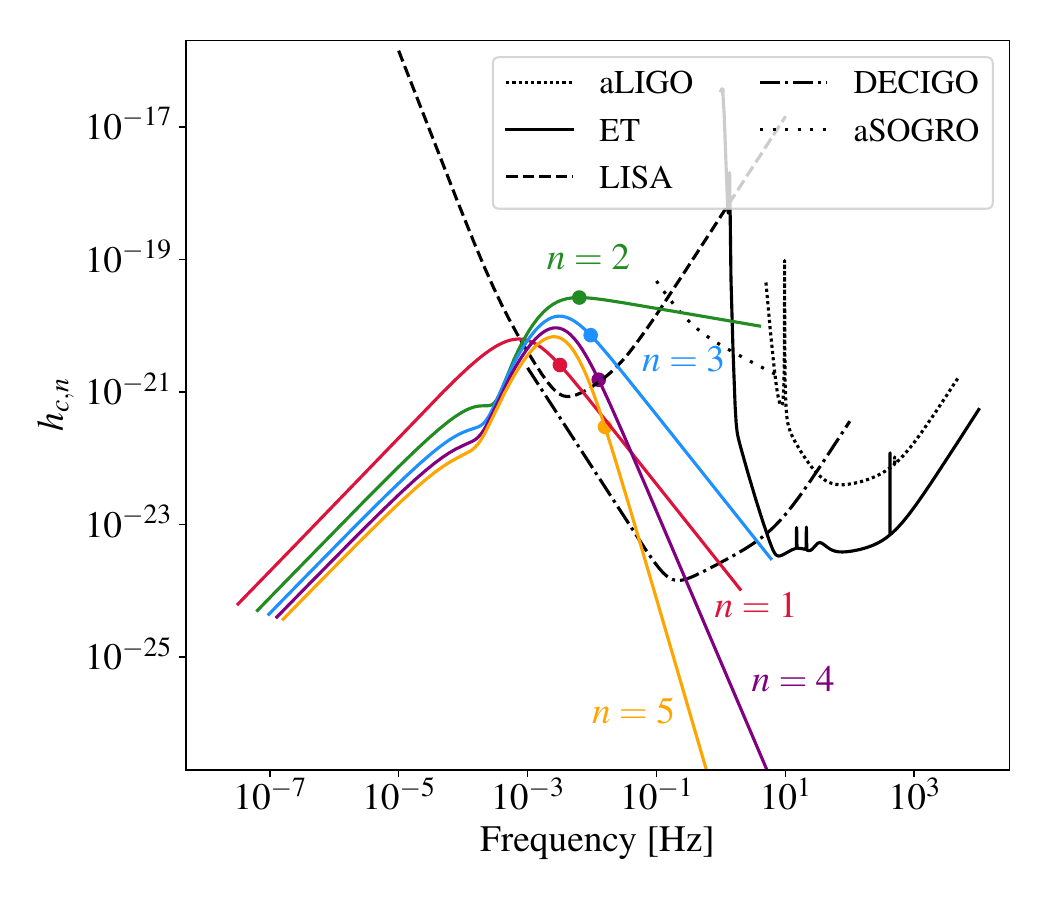}
    \vspace{-5mm}    
    \caption{A sample illustration of the characteristic amplitude of the $n$-th harmonics, $h_{c,n}$, from $n = 1$ to $5$ emitted during the IMRI. 
    The {\it dashed black V-shaped curve} represents LISA's sensitivity curve ($\sqrt{f S_h\left(f\right)}$; see Section \ref{sec: snr}). while two {\it U-shaped curves} indicate the sensitivities of aLIGO ({\it densely-dotted line}) and ET ({\it solid line}).
    Lastly, the intermediate frequency band from $10^{-3}\,\mathrm{Hz}$ to $10^1\,\mathrm{Hz}$ is covered by two detectors; aSOGRO ({\it loosely-dotted line}) and DECIGO ({\it dash-dotted line}).
    The dots on the curve mark the moment 4 years before the merger. 
    Here, we assume the IMBH mass to be $M_{\rm IMBH} = 10^3\msun$ and the secondary BH mass to be $M_{\rm BH} = 40\msun$. 
    The binary system is initialized with a semi-major axis $a=10\au$ and an eccentricity $e=0.999$, which is consistent with the green curve in Figure \ref{fig: Evolution of IMRI}. 
    The source is assumed to be located at a distance of $500\Mpc$ from us. 
    This figure demonstrates that the IMRI stays in the LISA and DECIGO detection band for most of its evolution.
    While the aSOGRO and the ET can capture the last stable merger, aLIGO is mostly insensitive to this GW event.}
    \label{fig: GW_Signals}
    \vspace{1mm}    
\end{figure}

\subsection{Prescriptions of Gravitational Wave (GW) Mergers and Tidal Disruptions} \label{sec: prescription}

This section briefly introduces how the code handles close encounters and then describes how we implement GW mergers and TDs for IMBHs in our simulations.
The IMBH embedded in a dense star cluster experiences multiple interactions with stars and BHs.
A close encounter between the two particles with masses $M_1$ and $M_2$ at a distance $R$ is treated by the KS regularization scheme,\footnote{KS regularization is a treatment introduced to reduce the computational load during close encounters between particles. This should be distinguished from the merger criteria of the binaries described in the following paragraphs.} provided the following conditions are met:
\begin{itemize}
    \item the two particles are closer than $r_{\rm min}$ (\texttt{RMIN}), and 
    \item their time-step becomes less than $\Delta t_{\rm min}$ (\texttt{DTMIN}).
\end{itemize}
Both $r_{\rm min}$ and $\Delta t_{\rm min}$ are user-controlled parameters in {\it N}-body units (See Appendix \ref{Appendix: unit}).
We consistently choose $r_{\rm min} = 10^{-4} r_N$ and $\Delta t_{\rm min} = 10^{-5} t_N$ across all simulations presented here, suppressing an auto-adjustment during the run.
Here, $r_N\approx 1.3 R_{\rm hm}$ and $t_N \approx 15 \Myr (r_N/M_{\rm Tot})^{1/2}$ are {\it N}-body length and time unit, respectively, where $R_{\rm hm}$ is the half-mass radius of the star cluster, and $M_{\rm Tot}$ is the initial total mass in our simulations.

Our GW merger and TD models operate by exmaning the orbital properties of all objects within $10^{-4}\pc$ when the IMBH is in a ``single'' (i.e., not undergoing any close encounter that meet the criteria for KS regularization), or by analyzing the orbital properties of the companion object when the IMBH is in a KS regularized state.
When the IMBH-BH binary undergoes a close encounter with other particles, their evolution is treated in the routine such as \texttt{triple}, \texttt{quad}, or \texttt{chain}.
In this case, however, we do not apply our model, as such events are rare. See Appendix \ref{Appendix: encounter timescale} for more justification.

If the IMBH is in a ``single'' state, we compute the semi-major axis $a$ and eccentricity $e$ of particles within a distance of $10^{-4}\pc$ at every time-step in \texttt{intgrt.f}.
If the IMBH forms a KS pair, we check the values of $a$ and $e$ for the binary during every call of the routine \texttt{ksint.f}.
We then decide whether to merge the particles based on the corresponding criteria regardless of KS regularization.
For a SBH, we calculate the merger timescale using the equation from \citet{Peters1964PhRv..136.1224P} as
\begin{equation}
    t_{\rm GW} = \frac{5}{256}\frac{c^5 a^4\left(1-e^2\right)^{3.5}}{G^3 \,M_{\rm IMBH} \,M_{\rm BH}\left(M_{\rm IMBH} + M_{\rm BH}\right)},
\label{eq: T_gw}
\end{equation}
where $M_{\rm IMBH}$ and $M_{\rm BH}$ refer to the masses of the IMBH and the SBH, respectively.
Then the binary merges instantly if $t_{\rm GW} < 1.0\,{\rm Myr}$ during the run.\footnote{See Appendix \ref{Appendix: hardness of binary} for more justification}
This timescale is sufficiently short to assume that the influence of the third particle on the binary is negligible. 
Therefore, we can safely consider that the binary evolves only through GW radiation.
After the merger, we increase the mass of the IMBH by $M_{\rm BH}$.

Although our primary goal is to investigate IMBH-BH collisions, implementing TD is necessary to properly consider and remove stars that are very close to the IMBH.
In the case of an IMBH-star encounter, we compute the pericenter of the orbit as $r_{\rm p} = a\left(1 - e\right)$.
Then, we remove the star from the simulation if $r_{\rm p}$ is less than the tidal radius, $r_{\rm T}$.
The tidal radius is computed by $r_{\rm T} = r_\star \left(M_{\rm IMBH}/M_\star\right)^{1/3}$, where the stellar radius computed by $r_{\star} = R_\odot\left(M_\star/M_\odot\right)^{1/3}$ for a star with mass $M_\star$.
In this case, the mass of the IMBH is increased by half of the removed star's mass, $0.5\, M_\star$.
The choice of accreting a half of the stellar mass is based on analytical estimates \citep[e.g.,][]{Rees1988Natur.333..523R, Strubbe2009MNRAS.400.2070S}.
The position and velocity of the IMBH are adjusted after a merger or a TD event in such a way that momentum conservation is guaranteed. 

\subsection{Signal-to-noise Ratios of Intermediate Mass Ratio Inspirals (IMRIs) at GW Observatories} \label{sec: snr}

The binary of BHs with masses $M_1$ and $M_2$ shrinks and circularizes by emitting energy as gravitational waves.
If external perturbations are negligible, we can analytically predict their evolution based on the 2.5PN (2.5 post-Newtonian) approximation.
We compute the evolution of the IMRIs based on the orbit-averaged prescription provided by \citet{Peters1964PhRv..136.1224P}.
Figure \ref{fig: Evolution of IMRI} depicts the evolution of the IMBH-BH binary with $a = 10\au$ with three different eccentricities, as calculated from Eqs. (\ref{eq: dadt}) and (\ref{eq: dedt}).
An IMRI with relatively modest eccentricity, $e = 0.9$, requires $\sim 10^{12}$ years to merge via pure GW radiation.
In contrast, it takes less than $10^5\yr$ for a highly eccentric IMRI until the merger.
The merger timescale of the binary is dramatically shorter for a more eccentric binary.

We then compute the sky-averaged signal-to-noise ratio (SNR) produced by the GW source at a luminosity distance $D_\mathrm{L}$ by combining the results from \citet{Finn2000PhRvD..62l4021F} and \citet{Barack2004PhRvD..69h2005B} as 
\begin{equation}
    \left(\frac{S}{N}\right)^2 = \sum_n\int{\frac{h_{c,n}^2\left(f_n\right)}{5 S_h\left(f_n\right)}}\frac{d\ln f_n}{f_n},
    \label{eq: S/N}
\end{equation}
where $f_n = n f = n\sqrt{G \left(M_1 + M_2\right)/a^3}$ is the frequencies of the $n$-th harmonic,
$h_{c,n}=\left(\pi D_{\rm L}\right)^{-1}\sqrt{2\dot{E}_n/\dot{f}_n}$ is the characteristic amplitude of the $n$-th harmonic at distance $D_{\rm L}$, $\dot{E}_n$ is the energy radiated through the $n$-th harmonic, and $S_{h}(f)$ represents the noise spectral density of the detectors at frequency $f$.\footnote{The prefactor 5 emerges as a result of averaging the noise over all-source directions \citep{Finn2000PhRvD..62l4021F}.}
We explore the signals onto five detectors; aLIGO, ET, LISA, aSOGRO, and DECIGO.
See Appendix \ref{Appendix: sensitivity} for more information. 
The integration domain of $f$ starts 4 years before the binary reaches the innermost stable circular orbit (ISCO).

Figure \ref{fig: GW_Signals} shows the evolution of the first five characteristic amplitudes, $h_{c,n}$ of an IMRI.
The IMRI considered consists of an IMBH with $M_{\rm IMBH} = 10^3 \msun$ and a BH with $M_{\rm BH} = 40\msun$.
The figure includes the curve of integrated GW signals and the intrinsic sensitivity curve for GW observatories.
The GW frequency increases as the binary shrinks, and it remains in the LISA band for a period of time.
The inspiral-plunge transition occurs approximately when the binary's orbital frequency crosses the ISCO frequency.
The frequency of its 2nd harmonic is given by
\begin{align}
    f_{2, \mathrm{ISCO}} &= \frac{1}{\pi(1 + z)}\sqrt{\frac{G M_{\rm IMBH}}{R^3_{\rm ISCO}}} \nonumber\\
    &\approx \frac{8.6}{1 + z}\,\mathrm{Hz} \left(\frac{M_{\rm IMBH}}{10^3\msun}\right)^{-1}.
    \label{eq: f_ISCO}
\end{align}
where $1/(1 + z)$ reflects the effect of the redshift due to the expansion of the universe.

\subsection{Modelling IMRI Detection Rates} \label{sec: IMRI Detection Rate}

\begin{figure}
    \centering
    \vspace{-3mm}
    \includegraphics[width=1.0\columnwidth]{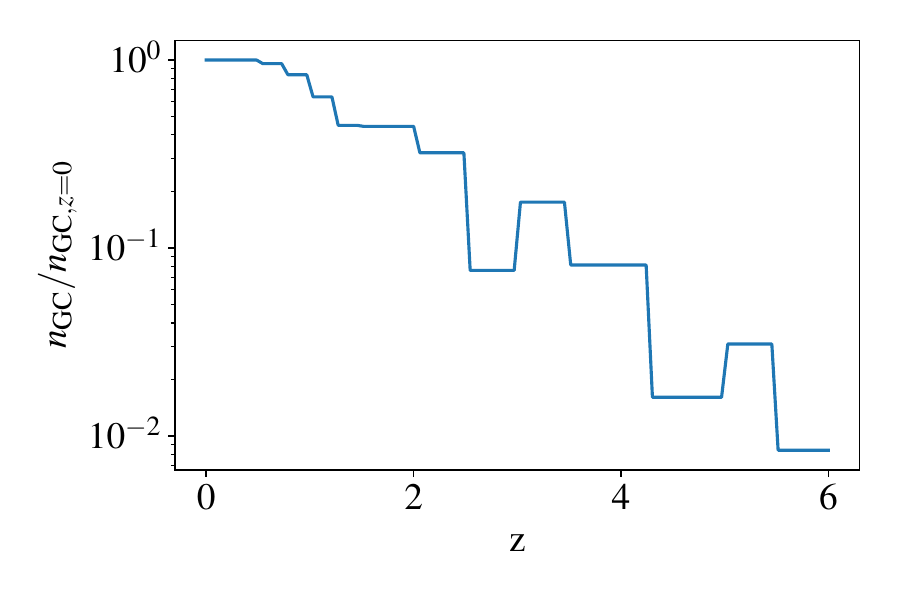}
    \caption{The evolution of the ``normalized'' GC number density, $\nu_{\rm GC} \equiv n_{\rm GC}/n_{{\rm GC}, \,z = 0}$, from $z = 0$ to $z = 6$.
    The local GC number density $n_{{\rm GC}, \,z = 0}$ is estimated based on the quantities such as the galaxy populations and the average fraction of galactic masses contained in GCs.
    The rebounds at $z = 3$, and $z = 4.5$ arise from the discords between different measurements; we choose \citep{Tomczak2014ApJ...783...85T} for $z = 0 - 3$, \citep{Caputi2011MNRAS.413..162C} for $z = 3 - 4.5$, and \citep{Song2016ApJ...825....5S} for $z = 4.5 - 6$.
    See Section \ref{sec: IMRI Detection Rate} for more information.}
    \label{fig:n_GC}
    \vspace{1mm}    
\end{figure}

Due to the limited computational resources, our star clusters should have smaller particle numbers and sizes than the realistic star clusters.
Therefore, we try to mimic the GCs by analyzing the IMRI rates as a function of velocity dispersion given by the following form:
\begin{equation}
    \Gamma_{\rm IMRI} \propto M_{\rm IMBH}^A \sigma_{\rm hm}^B,
    \label{eq: Linear_Fit_IMRI}
\end{equation}
to establish an empirical relation estimating the IMRI rates (see Eq.(\ref{linear_fit}) and Section \ref{sec: IMRI Statistics}).
Here, $\sigma_{\rm hm}$ is the 1D half-mass velocity dispersion of the cluster.

Our simulations also provide statistics on the orbital properties of IMRIs, such as the eccentricities or the masses of secondary BHs.
We then estimate the detection rate of these events using the formula:
\begin{align}
    & \frac{\mathrm{d} \Gamma_{\rm det}}{\mathrm{d} M_{\rm GC}\,\mathrm{d} M_{\rm IMBH}\,\mathrm{d} z} \nonumber\\
    & = p_{\rm IMBH} f_{\rm det}\left(M_{\rm IMBH}, z\right) \Gamma_{\rm IMRI}\left(M_{\rm IMBH}, \sigma_{\rm hm}\right) \nonumber\\
    & \,\,\,\,\,\,\,\, \times n_{\rm GC} \frac{\mathrm{d} \nu_{\rm GC}}{\mathrm{d} M_{\rm GC}} \frac{\mathrm{d} M_{\rm GC}}{\mathrm{d} M_{\rm IMBH}} \frac{\mathrm{d} V_c}{dz}\frac{1}{\left(1+z\right)},
\end{align}
which contains the following components:
\begin{itemize}
    \item $p_{\rm IMBH}$ is the IMBH retention probability of GCs. 
    The magnitude of the natal kick can be modeled as a Maxwellian distribution \citep{Hobbs2005MNRAS.360..974H}, or it can be derived from the mass fallback caused by the supernova explosion during the IMBH synthesis \citep{Belczynski2002ApJ...572..407B}.
    About $\lesssim 30\%$ of GCs are expected to host IMBHs, and this probability may fall close to $0\%$ for GCs with masses below $10^5\msun$ \citep{Giersz2015MNRAS.454.3150G, Askar2017MNRAS.464L..36A, Askar2022MNRAS.511.2631A}.
    We utilize the retention probability of IMBH formation depending on the star cluster mass considering both high and low natal kicks referring Table 1 in \citet{Askar2022MNRAS.511.2631A}.
    \item $f_{\rm det}$ is the fraction of IMRIs observable by a GW detector, which depends on $M_{\rm IMBH}$, redshift $z$, and the specific detector design.
    We assume that detectable IMRIs must have a signal-to-noise ratio $S/N > 8$. 
    This factor is modeled based on the distribution of orbital properties of IMRI events derived from our simulations.
    \item $\mathrm{d}V_c /\mathrm{d}z$ is the comoving volume element at a given redshift bin, and $1/(1+z)$ accounts for the time dilation.        
    \item $n_{\rm GC}$ is the number density of GCs in each redshift bin.
    Figure \ref{fig:n_GC} presents the ``normalized'' GC number density, $\nu_{\rm GC} \equiv n_{\rm GC}/n_{{\rm GC}, \,z = 0}$, at each redshift bin, while the absolute value of $n_{\rm GC}$ depends on the estimation of local GC density.
    Indeed, the GC number density at $z = 0$ varies across studies --- from $0.77\Mpc^{-3}$ \citep{Rodriguez2015PhRvL.115e1101R} to $8.4\,h^3\Mpc^{-3}$ \citep{Zwart1999astro.ph.12022P}.
    In our approximation, assuming GCs with masses $5\times 10^4$ to $8\times 10^6 \msun$ account for $0.1\%$ of the galaxy mass, we obtain $n_{{\rm GC}, \,z = 0} = 1.45 \Mpc^{-3}$.
    Note that the average mass of these GCs becomes $\left<M_{\rm GC}\right> = 1.9\times 10^5\msun$ here.
    We also model the halo mass function, $\phi(\Bar{M},z)$, by the \citet{Schechter1976ApJ...203..297S}  function (see Appendix \ref{Appendix: schechter} for detailed information), with the parameters adopted from Table.(1) of \citet{Conselice2016ApJ...830...83C}.    
    More details about $n_{\rm GC}$ can be found in Appendix \ref{Appendix: n_GC}.
    \item $\mathrm{d} \nu_{\rm GC}/\mathrm{d} M_{\rm GC}$ refers to the normalized GC mass function. 
    We assume that it follows a simple power-law, $\mathrm{d} \nu_{\rm GC}/\mathrm{d} M_{\rm GC} \propto M_{\rm GC}^{-s}$ with a slope parameter $s = 2.2$ \citep{Gieles2009MNRAS.394.2113G}.
    \item $\mathrm{d}M_{\rm GC}/\mathrm{d} M_{\rm IMBH}$ models the relation between $M_{\rm IMBH}$ and $M_{\rm GC}$. 
    To simplify our calculation, here we use the linear relation from \citet{Arca-Sedda2016MNRAS.455...35A},
    \begin{equation}
        \log{\left(\frac{M_{\rm IMBH}}{\msun}\right)} = \log{\left(\frac{M_{\rm GC}}{\msun}\right) - 2.23}.
        \label{eq: IMBH-GC relation}
    \end{equation}
    According to this relation, IMBHs of $300 - 5000\msun$ correspond to the GCs of masses  $5\times 10^4 - 10^6\msun$.
\end{itemize}

%\vspace{1mm}

\begin{figure}
    \centering
    \vspace{-3mm}
    \includegraphics[width=0.93\columnwidth]{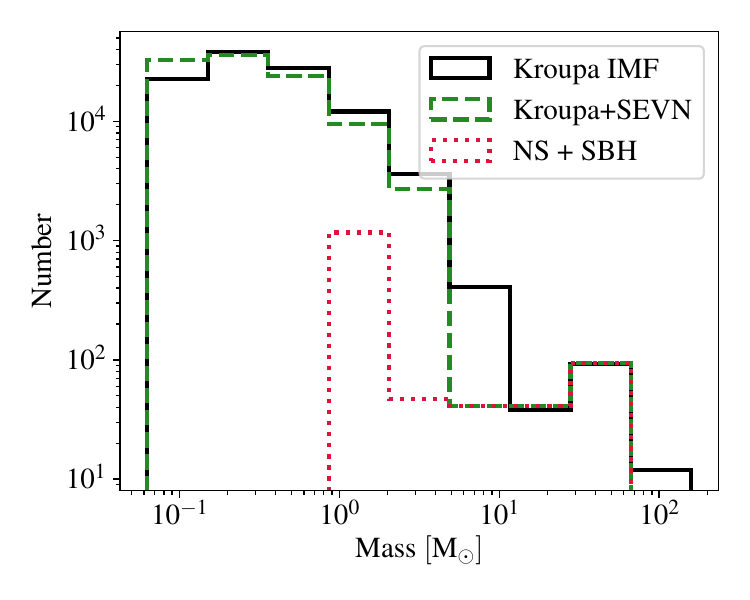}
    \caption{The distribution of stellar populations sampled from the Kroupa IMF ({\it solid-black}) with mass ranges $0.08 - 150 \msun$.
    The population after evolving $100\Myr$ loses some of its heavy components and the maximal mass decreases to $60\msun$ due to the mass loss during evolution ({\it dashed-green}).
    Especially, compact objects ({\it dotted-red}) such as stellar-mass black holes (SBHs) or neutron stars (NSs) that are subject to GW merger are evenly distributed over the mass ranges $0.8 - 60\msun$.}
    \label{fig: IMF}
    \vspace{1mm}    
\end{figure}

\begin{table}
    \caption{Initial conditions of our simulation suite. 
    Both Groups A and B investigate the influence of the star cluster properties and the IMBH mass on the GW merger rate, respectively.
    Each simulation has been tested over 10 random seeds.
    The initial half-mass velocity dispersion is calculated for particles within $R_{\rm hm}$ from the IMBH after the ``adiabatic" insertion.
    Although the star clusters are initially generated by identical properties (Group B), the velocity dispersion is affected by the mass of IMBHs.}
    \vspace{2mm}
    \centering
    \begin{tabular}{lcccccc}
        \hline
        Name{\textdagger} & $M_{\rm GC}$ & $N_{\rm star}$ & $R_{\rm hm}$ & $\sigma_{\rm hm}$ & $M_{{\rm IMBH},i}$ & $t_{\rm term}$\\
        & $\left(\msun\right)$ & $\left(10^4\right)$ & $\left(\pc\right)$ & $\left(\km\s^{-1}\right)$ & $\left(\msun\right)$ & $(\Myr)$\\
        \hline
        \hline
        A-i &  $5\times 10^4$ & 10.5 & 0.5 & 8.52 & $1000$ & 30 \\
        A-ii\footnote{Identical to B-iii} & $5\times 10^4$ & 10.5 & 0.7 & 7.27 & $1000$ & 30 \\
        A-iii &  $5\times 10^4$ & 10.5 & 1.0 & 6.07 & $1000$ & 30 \\
        A-iv &  $10^5$ & 21.0 & 0.7 & 10.06 & $1000$ & 30 \\
        A-v &  $10^5$ & 21.0 & 1.0 & 8.41 & $1000$ & 30 \\
        \hline
        B-i & $5\times 10^4$ & 10.5 & 0.7 & 7.06 & $300$ & 30 \\
        B-ii & $5\times 10^4$ & 10.5 & 0.7 & 7.15 & $500$ & 30 \\
        B-iii & $5\times 10^4$ & 10.5 & 0.7 & 7.27 & $1000$ & 30 \\
        B-iv & $5\times 10^4$ & 10.5 & 0.7 & 7.56 & $2000$ & 30 \\ 
        B-v & $5\times 10^4$ & 10.5 & 0.7 & 7.84 & $3000$ & 30 \\
        B-vi & $5\times 10^4$ & 10.5 & 0.7 & 8.30 & $5000$ & 30 \\
        \hline
    \end{tabular}
    \tablecomments{\scriptsize \,\,\,{\textdagger}\,The list of simulations performed for this study. Col. 1: Names of simulations. Col. 2-3: Total mass and total number of particles in the star cluster. Col. 4-5: Half-mass radius and initial 1D half-mass velocity dispersion of the cluster. Col. 6: Initial mass of an embedded IMBH. Col. 7: Termination time of the simulation.}
    \vspace{3mm}
    \label{tab: Initial Conditions}
\end{table}

\section{Initial Conditions} \label{sec: Initial Conditions}

\begin{figure}
    \centering
    \vspace{2mm}
    \includegraphics[width = 0.93\columnwidth]{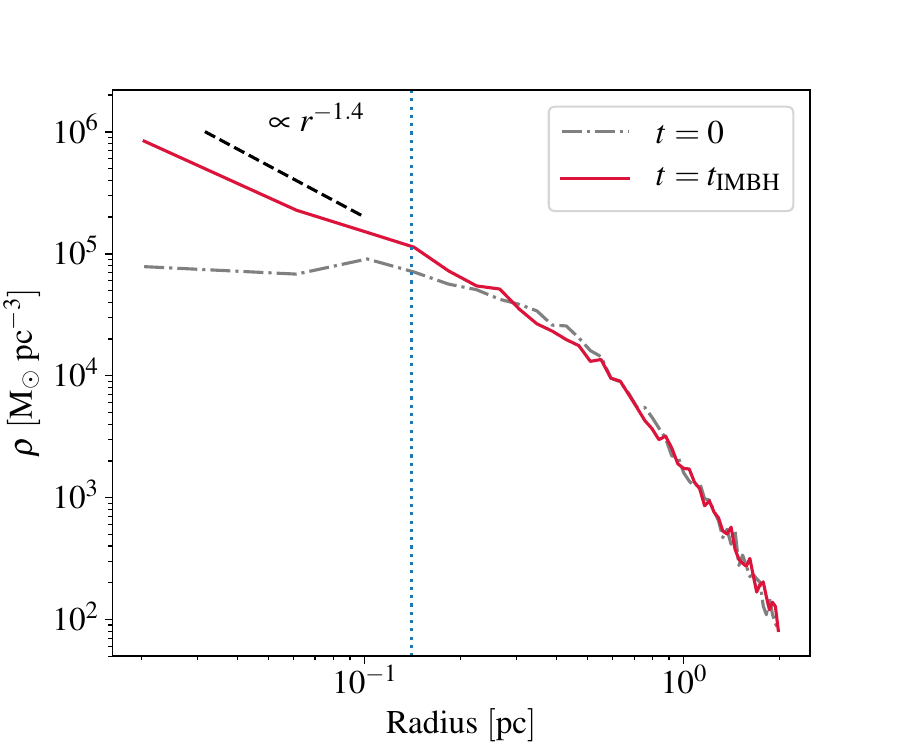}
    \vspace{-1mm}    
    \caption{The radial density profile of the GC before ({\it grey} {\it dash-dotted}) and after ({\it red}) the ``adiabatic growth'' of a star cluster to introduce an IMBH (see Section \ref{sec: adiabatic growth} for more information).
    The star cluster has a mass of $5\times 10^4 \msun$ and a half-mass radius of $0.7 \pc$. 
    The IMBH mass when it is introduced as a live particle is $M_{{\rm IMBH}, i}= 10^3\msun$. 
    The inner profile of the star cluster is flat at $t = 0$, whereas it is deformed to follow a power-law profile at $t = t_{\rm IMBH}$ ($= 50\, t_{\rm N}$; see Section \ref{sec: adiabatic growth}). 
    The simulated star cluster has a slope of $\gamma \approx 1.4$, less steep than the theoretically expected value of $\gamma = 1.75$. 
    The {\it vertical dotted line} refers to the IMBH's radius of gravitational influence.}
    \label{fig: Adiabatic Insert}
    \vspace{2mm}
\end{figure}

\subsection{Star Cluster Parameters and Computational Optimizations}\label{sec: sc parameters}

We generate the initial conditions of the star cluster using the code {\sc Mcluster} \citep{Kupper2011MNRAS.417.2300K}. 
The simulated star clusters are designed to follow the Plummer profile,
\begin{equation}
    \rho_{\rm P}\left(r\right) = \frac{3 M_{\rm SC}}{4\pi a_{\rm P}^3}\left(1 + \frac{r^2}{a_{\rm P}^2}\right)^{-2.5},
\end{equation}
where $M_{\rm SC}$ and $a_{\rm P}$ refer to the total mass and scale size of the star cluster, respectively.
The half-mass radius $R_{\rm hm}$ of the Plummer cluster is approximated by $1.3 a_{\rm P}$.
We initialize the clusters without including primordial mass segregation or binaries.

To construct a realistic SBH mass function, we generate the stellar population by sampling from the \citet{Kroupa2001MNRAS.322..231K} initial mass function (IMF) within stellar mass range of $0.08 - 150\msun$. 
We then evolve the stars using the stellar evolution code \texttt{SEVN} \citep{Spera2019MNRAS.485..889S, Iorio2023MNRAS.524..426I} over $100\Myr$.
Figure \ref{fig: IMF} presents the mass function before and after stellar evolution.
It reveals that the neutron stars (NSs) and SBHs are spread over the mass range from $0.8\msun$ to $60\msun$.
Most stars with initial masses above $10 \msun$ turn into the BHs, with their final mass decreasing by the effect of the pair-instability supernova (PISN) or pulsational pair-instability supernove (PPISN) \citep[e.g.,][]{Woosley2017ApJ...836..244W, Woosley2021ApJ...912L..31W}.
Meanwhile, the NSs dominate the lower-mass end of the compact object spectrum.
Since our simulations run over a relatively short timescale up to $30\Myr$, we suppress further stellar evolution during our experiments.
This is because stellar evolution processes extend over significantly longer timescales and are beyond the primary focus of our study.

The typical mass range of GCs is a few$\times 10^3$ to $5\times 10^6\msun$ with a scale radius of $\sim 3\pc$.
However, higher-mass GCs still need $\gtrsim \mathcal{O}\left(10^5\right)$ particles to be simulated.
We therefore mimic the dense environment of GCs by reducing the size of GCs instead of increasing their mass.
This approach allows us to maintain a manageable number of particles resolved in simulations, thus reducing computational costs.
We note that the overall evolution of the star cluster is significantly influenced by its total mass.
However, the effect of the different GC masses on the GC evolution can be marginalized, as we terminate our simulations after only $t_{\rm term} = 30\Myr$ (more discussion in Section \ref{sec: Evolution of the star cluster}).
%This helps to avoid the large deviations caused by differences in GC evolution over longer periods.

We explore the IMRI rates in various stellar system settings, by changing its initial half-mass velocity dispersion from $\sigma_{\rm hm} \sim 6\,\mathrm{km}\,\mathrm{s}^{-1}$ to $\sim 10\,\mathrm{km}\,\mathrm{s}^{-1}$ (Group A in Table \ref{tab: Initial Conditions}), and by varying the initial IMBH mass from $M_{{\rm IMBH}, i} = 300\msun$ to $5000\msun$ (Group B in Table \ref{tab: Initial Conditions}).
Before inserting the IMBH, we gradually modify the GC centers by slowly increasing the point mass potential, in order to avoid the sudden introduction of a massive object (as described in Section \ref{sec: adiabatic growth}).
All our simulated star clusters are initialized in isolation, with no external tidal fields included.

\subsection{Adiabatic Growth of A Star Cluster To Introduce An Embedded IMBH}\label{sec: adiabatic growth}

\begin{figure}
    \centering
    \includegraphics[width=0.93\columnwidth]{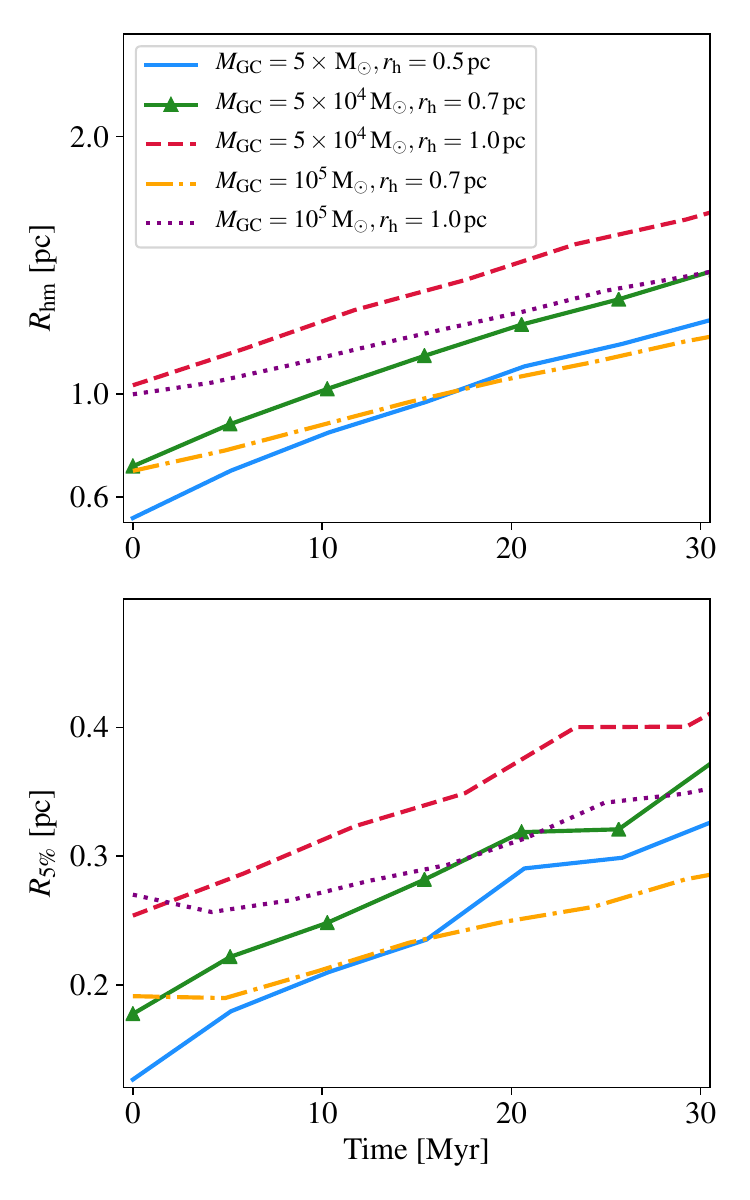}
    \vspace{-5mm}   
    \caption{This plot depicts the evolution half-mass radii ({\it top}) and the radius enclosing 5\% of the initial cluster mass ({\it bottom}) for Group A.
    Each curve is averaged over all 10 random seeds, and the mass of the IMBH is excluded from the calculations. The overall increase of the radius shows the expansion of the stellar systems.
    }
    \label{fig: LagR_Group_A}
    \vspace{0mm}
\end{figure}

\begin{figure}
    \centering
    \includegraphics[width=0.93\columnwidth]{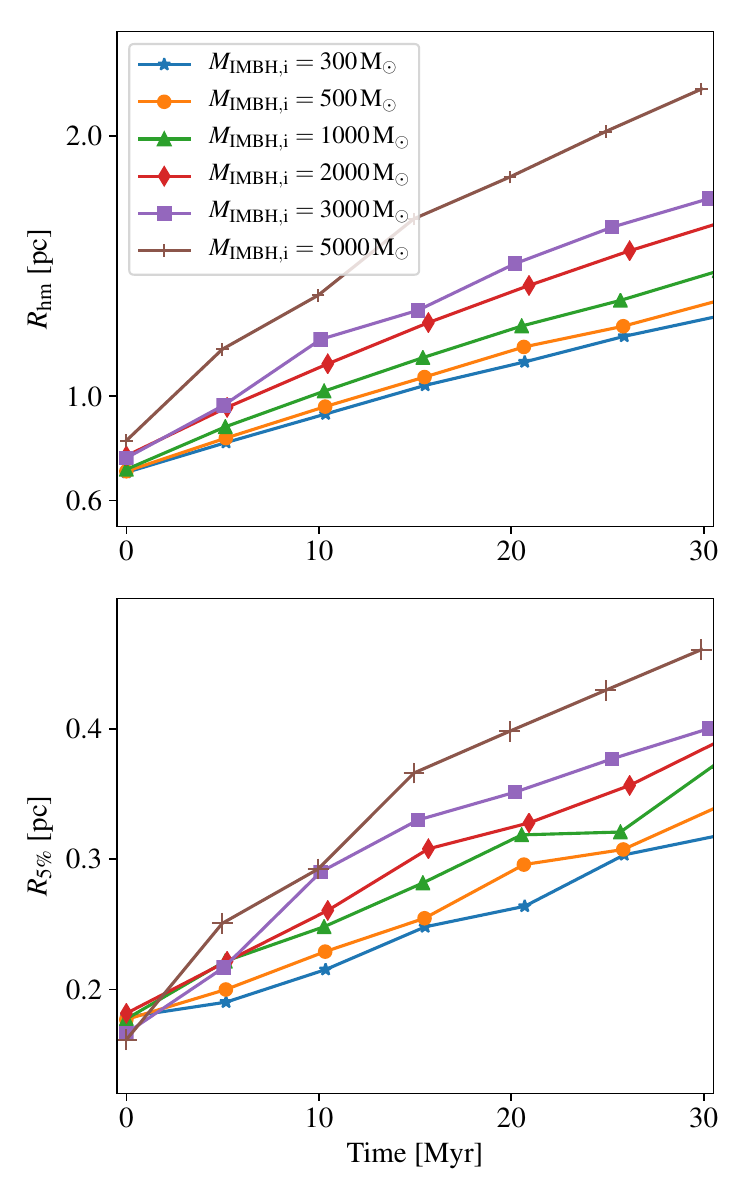}
    \vspace{-4mm}   
    \caption{The same set of plots as presented in Figure \ref{fig: LagR_Group_A}, but now corresponding to Group B.}
    \label{fig: LagR_Group_B}
    \vspace{0mm}    
\end{figure}

\begin{figure*}
    \centering
    \vspace{-2mm}     
    \includegraphics[width = 0.85\textwidth]{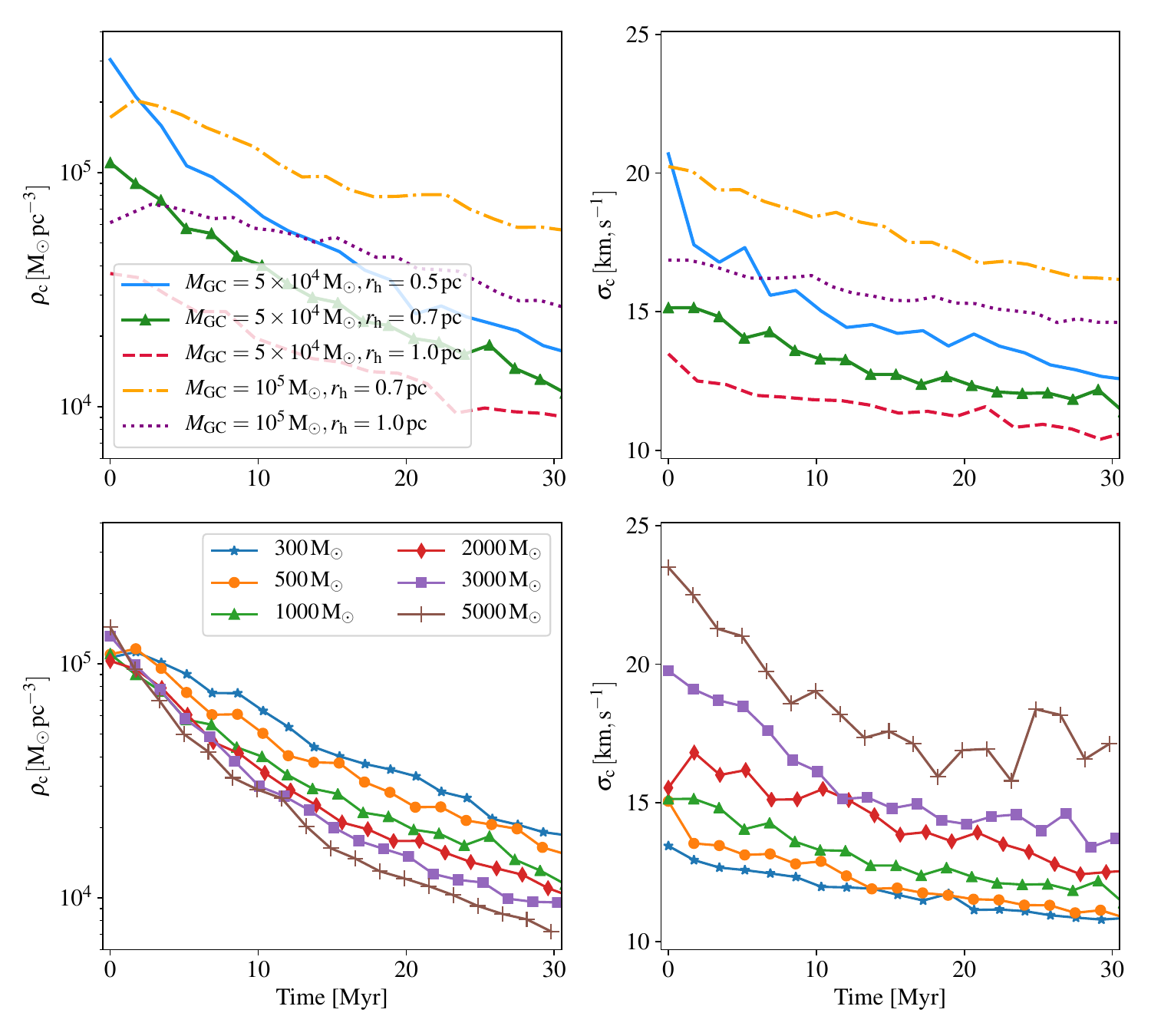}
    \vspace{-2mm}    
    \caption{The evolution of the mean density ({\it left}), and the (1D) velocity dispersion ({\it right}) near the cluster center for both group A ({\it top}) and Group B ({\it bottom}). 
    These quantities are computed over the population except for the IMBH within $r < R_{5\%}$ from the center.
    In general, the cluster density and velocity dispersion decrease over time due to the expansion of the star clusters.
    Both the density and velocity dispersion within $r < R_{5\%}$ are significantly affected by the initial cluster properties while it is less affected by the initial mass of the IMBH.
    In Group A, the variations in density and velocity dispersion across different runs suggest that the initial parameters of the star clusters play a dominant role in determining the long-term evolution of the cluster core.
    In Group B, the more rapid decrease of density for higher-mass IMBH demonstrates the faster expansion of the star clusters.
    For an IMBH mass of $M_{\rm IMBH,i} = 5000\msun$, the initial velocity dispersion of the stars near the cluster center reaches $\sim 24\km \s^{-1}$.
    In contrast, for lower-mass IMBHs, such as $M_{\rm IMBH,i} = 300\msun$, the initial velocity dispersion is significantly lower, reaching only $\sim 14\km \s^{-1}$.
    See Section \ref{sec: Evolution of the star cluster} for more information.}
    \label{fig: Core Evolution}
    \vspace{2mm}
\end{figure*}

\begin{figure}
    \centering
    \includegraphics[width=0.93\columnwidth]{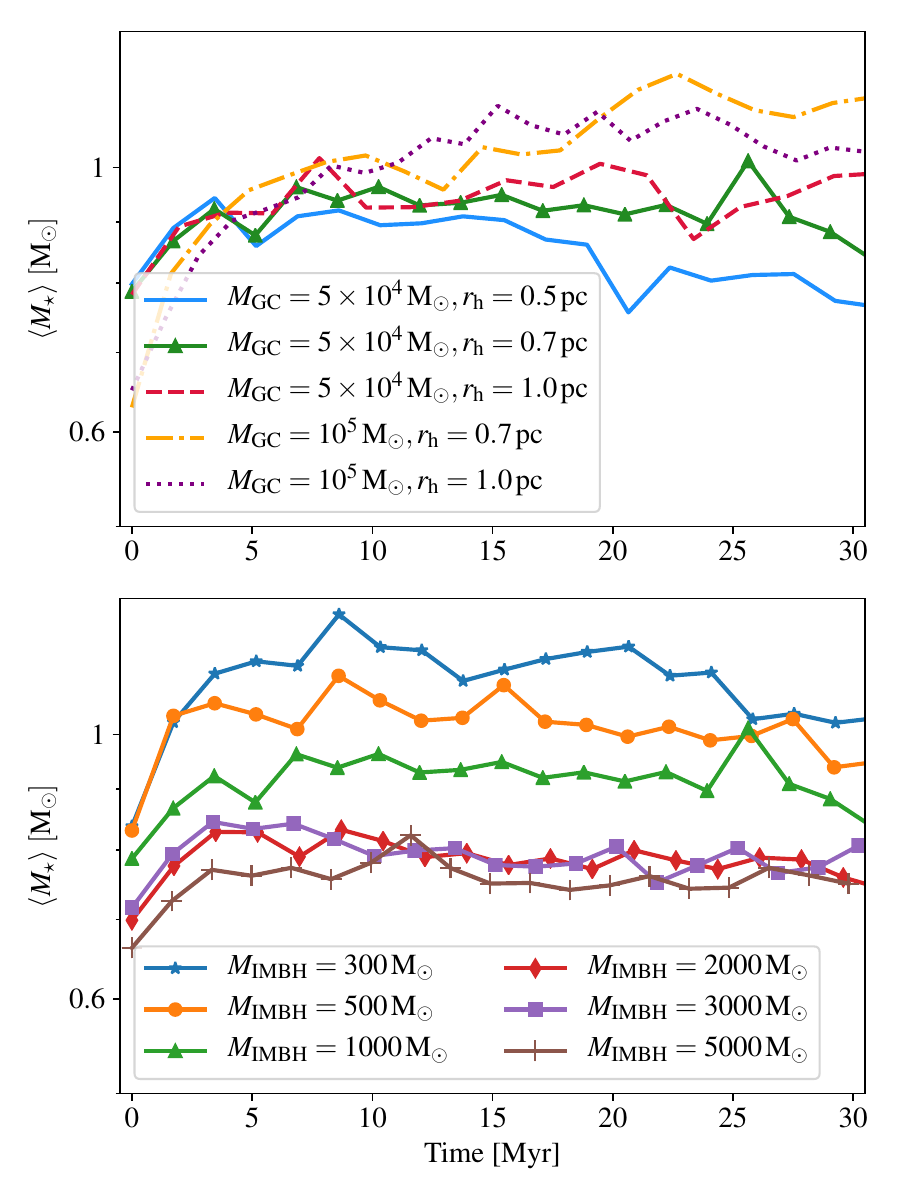}
    \vspace{-4mm}   
    \caption{The figure illustrates the evolution of the mean mass, $\left<M_\star \right>$, of stars and SBHs within a radius of $r < R_{5\%}$ presented separately for Group A ({\it top}) and Group B ({\it bottom}).
    In general, $\left<M_\star \right>$ exhibits a rapid increase in the early stage of evolution, stabilizing by $t = 10\Myr$ across all simulations.
    In the top panel, $\left<M_\star \right>$ gradually diverges over time, reflecting the influence of the initial properties of the GCs on their long-term evolution.
    Conversely, the bottom panel shows a decreasing trend in $\left<M_\star \right>$ with increasing $M_{\rm IMBH}$, suggesting that the presence of a more massive IMBH influences the retention and distribution of stellar populations within the core region. 
    The initial discrepancy at $t = 0$ is attributed to the evolution of star clusters during the adiabatic growth of IMBH.}
    \label{fig: Mean_Mass}
    \vspace{0mm}    
\end{figure}

\begin{figure*}
    \centering
    \vspace{-2mm}
    \includegraphics[width = 0.85\textwidth]{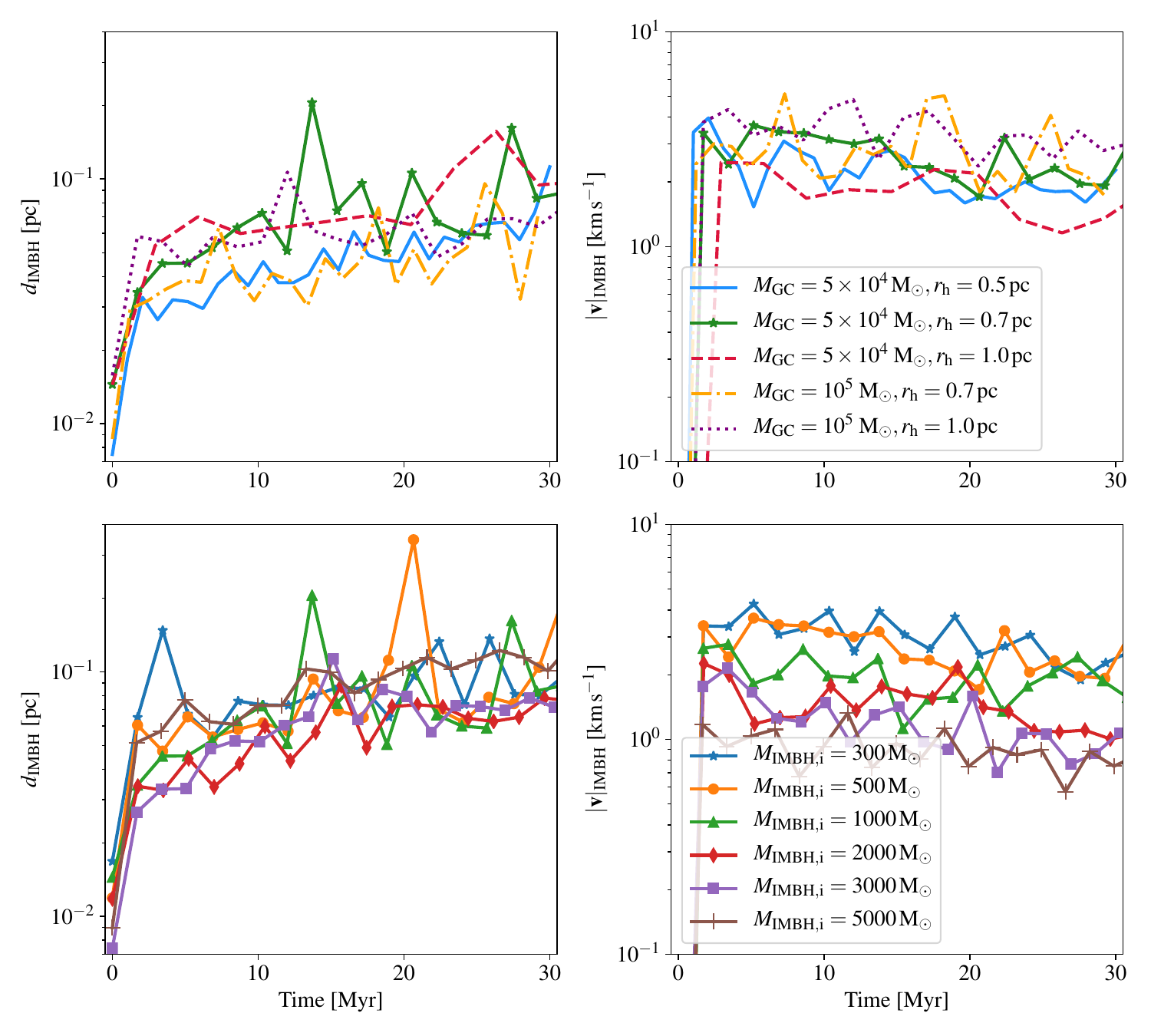}
    \vspace{-2mm}    
    \caption{The evolution of the IMBH's distance ({\it left}) from the center of mass of the star cluster and IMBH's speed ({\it right}), considering Group A ({\it top}) and Group B ({\it bottom}).
    The IMBH is slightly off ($\lesssim 0.1\pc$) from the center of the GC both in Group A and Group B.
    The dynamics of IMBHs show a weak dependence on the cluster properties in Group A, while they show a clear dependence on $M_{\rm IMBH}$ in Group B.
    Although the overall trend is well predicted by Equation \ref{eq: v_IMBH}, its scale is approximately an order of magnitude greater than the estimation.}
    \label{fig: IMBH_Dynamics}
    \vspace{2mm}
\end{figure*}

Both observational \citep[e.g.,][]{Schodel2009A&A...502...91S} and theoretical \citep[e.g.,][]{Bahcall1976ApJ...209..214B} studies agree that star clusters with a massive central object exhibit steeply increasing profiles for density and velocity dispersion toward the center.
\citet{Sigurdsson1995ApJ...446...75S} suggested that an ``adiabatically growing'' density profile can be a reasonable way to realize an $N$-body system with an IMBH.
First, before inserting the IMBH as a live particle, we generate the initial condition of star clusters following a Plummer density profile using {\sc Mcluster}. 
Then we introduce the Plummer potential with a gradually increasing mass $M_{\rm IMBH}\left(t\right)$ as
\begin{equation}
    \phi_{\rm IMBH} = - \frac{G M_{\rm IMBH}\left(t\right)}{\left(r^2 + \epsilon^2_{\rm IMBH}\right)^{1/2}},
\end{equation}
where $\epsilon_{\rm IMBH} \sim 10^{-4}\pc$ is the softening parameter placed to avoid the effect of the singularity \citep{Hong2015MNRAS.448..754H}.
Then, $t_{\rm IMBH}$ after the initialization, we replace the external potential with a live, moving IMBH particle.
Here, the time-dependent IMBH mass is designed to grow over time as
\begin{equation}
    M_{\rm IMBH}\left(t\right) = M_{{\rm IMBH}, i}\left[3\left(\frac{t}{t_{\rm IMBH}}\right)^2 - 2\left(\frac{t}{t_{\rm IMBH}}\right)^3\right],
\end{equation}
where $M_{{\rm IMBH}, i}$ is the initial mass of the IMBH when it is finally inserted as a live particle, and $t_{\rm IMBH}$ is the mass growth timescale.
Any choice of $t_{\rm IMBH}$ is suitable as long as it is not too short \citep{Sigurdsson1995ApJ...446...75S}, thus we set it to be $t_{\rm IMBH} = 50 t_{\rm N}$.
This is a relatively short timescale in our simulations, $\sim$ a few Myrs. 
The definition of the {\it N}-body units can be found in Appendix \ref{Appendix: unit}.
Figure \ref{fig: Adiabatic Insert} shows an example of ``adiabatic growth'' of a star cluster to introduce an embedded IMBH.
At $t = 0$, the cluster core has a flat density profile.
As the analytic IMBH potential gradually grows, the inner profile evolves into a cuspy profile of $\rho \propto r^{-\gamma}$ with a slope index $\gamma \approx 1.4$.
This is slightly shallower than the classical prediction by \citet{Bahcall1976ApJ...209..214B}, $\gamma = 1.75$.
This difference can be attributed to the nature of the star clusters consisting of stars with different masses \citep{Baumgardt2004ApJ...613.1143B}.

\begin{figure}
    \centering
    \includegraphics[width = 0.91\columnwidth]{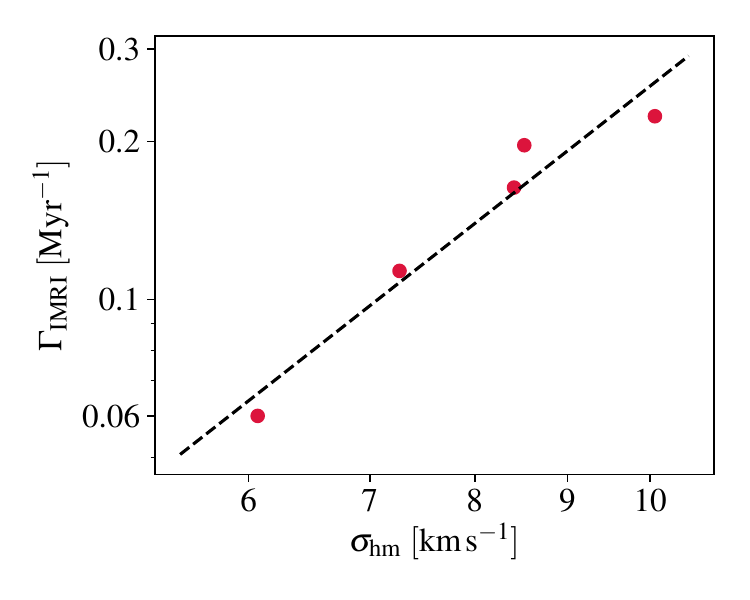}
    \vspace{-7mm}        
    \caption{The IMRI event rates, $\Gamma_{\rm IMRI}$, for an IMBH of initial mass $10^3\msun$ embedded in the GCs with initial half-mass velocity dispersion $\sigma_{\rm hm} \sim 6\,\mathrm{km}\,\mathrm{s}^{-1}$ to $\sim 10\,\mathrm{km}\,\mathrm{s}^{-1}$ (Group B in Table \ref{tab: Initial Conditions}). 
    Each data point is averaged over 10 random seeds.
    The {\it black dashed line} is the linear fit, indicating a clear positive correlation between $\sigma_{\rm hm}$ and $\Gamma_{\rm IMRI}$.
    See Section \ref{sec: IMRI Statistics} for more information.}
    \label{fig: IMRI Rate}
    \vspace{0mm}    
\end{figure}

\begin{figure}
    \centering
    \includegraphics[width=0.91\columnwidth]{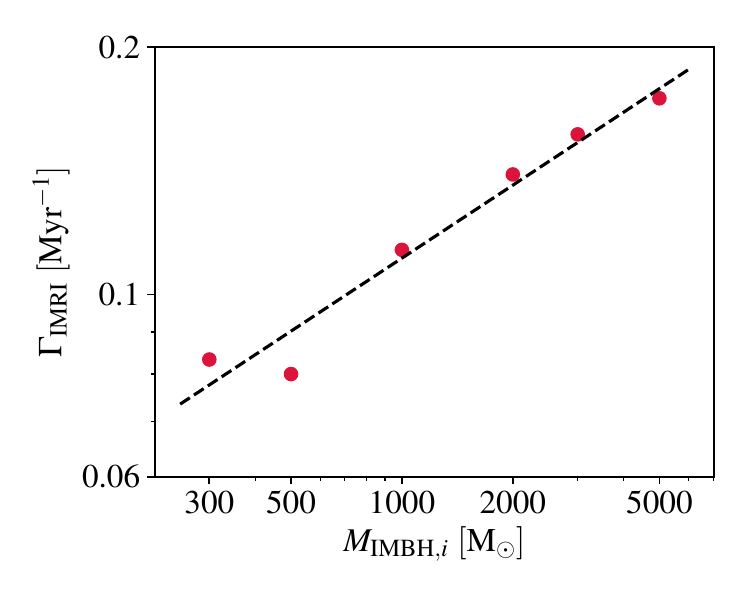}
    \vspace{-7mm}          
    \caption{The IMRI event rates, $\Gamma_{\rm IMRI}$, for an initial IMBH of mass $300$ to $5000\msun$ embedded in the GCs with a mass $5\times 10^4\msun$ and a half-mass radius $0.7\pc$ (Group C in Table \ref{tab: Initial Conditions}). 
    Each data point is averaged over 10 random seeds.    
    The {\it black dashed line} is the linear fit, indicating a positive correlation between $M_{{\rm IMBH}, i}$ and $\Gamma_{\rm IMRI}$.
    See Section \ref{sec: IMRI Statistics} for more information.}    
    \label{fig: IMRI_Rate_MBH}
    \vspace{0mm}    
\end{figure}

\begin{figure*}
    \centering
    \includegraphics[width = 0.98\textwidth]{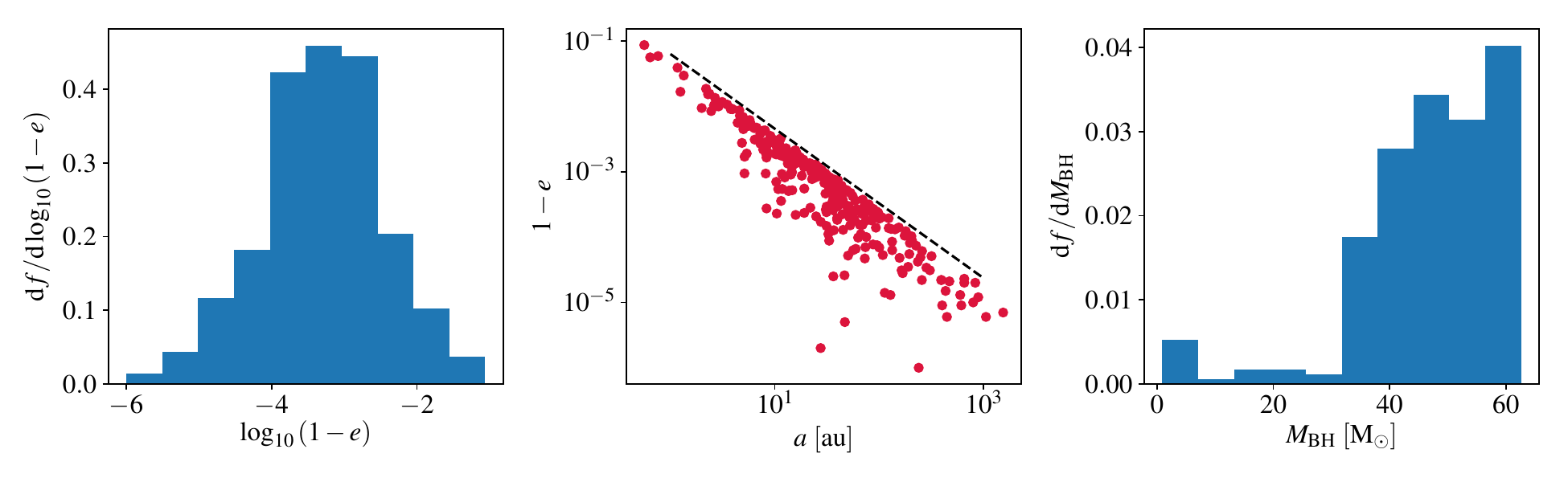}
    \vspace{-4mm}    
    \caption{The distribution of the orbital parameters of all the IMRI events in all simulations of Group A (see Table \ref{tab: Initial Conditions}). 
    The {\it left panel} shows the distribution of eccentricities, peaking at $\left(1 - e\right) \sim 10^{-3}$. 
    The {\it middle panel} displays the scatter plot of the semi-major axes $a$ versus eccentricities $e$, showing that a large number of eccentric binaries are formed by the IMBH.  
    The {\it black dashed line} indicates a boundary on which the merger timescale $t_{\rm GW}$ (Eq.(\ref{eq: T_gw}) in  Section \ref{sec: prescription}) is $1.0\Myr$ for $M_{\rm IMBH} = 10^3\msun$ and $M_{\rm BH} = 60\msun$. 
    Since we imposed that only the binaries with $t_{\rm GW} < 1.0\,{\rm Myr}$ are merged in the simulation (see Section \ref{sec: prescription}), most of our data points are located below this line, implying shorter merger timescales than $1.0\Myr$. 
    The {\it right panel} shows the distribution of the mass of the secondary, $M_{\rm BH}$, in the IMRI event, illustrating that more massive BHs are likely to merge with the IMBH.
    See Section \ref{sec: IMRI Statistics} for more information.}        
    \label{fig: Encounters}
    \vspace{-1mm}
\end{figure*}

\vspace{1mm}

\section{Results} \label{sec:Results}
Now we present the results of our simulations and discuss the possibility of detecting IMRIs through each of the existing and future GW telescopes.

\subsection{Evolution of A Star Cluster System}\label{sec: Evolution of the star cluster}

We begin by examining the temporal evolution of the star clusters.
The characteristic timescale for the evolution can be estimated by the relaxation timescale, $t_{\rm rlx}$, which is defined as
\begin{equation}
    t_{\rm rlx} = \frac{\sqrt{2} \sigma^3(r)}{\pi G^2 \left<M_\star\right>\rho(r)\ln {\lambda}}
    \label{eq: t_rlx}
\end{equation}
where $\sigma\left(r\right)$ is the local velocity dispersion, $\left<M_\star\right>$ is the average stellar mass, $\rho\left(r\right)$ is the mass density, and $\ln \lambda \approx 10$ is the Coulomb logarithm.
This expression suggests that denser star clusters evolve more rapidly compared to less dense clusters.

The evolution of the half-mass radius, $R_{\rm hm}$, and the Lagrangian radius enclosing 5\% of the initial cluster mass, $R_{5\%}$ of runs in Group B is depicted in Figure \ref{fig: LagR_Group_A}.
These radii serve as key indicators of the structural evolution of star clusters over time.
In general, the star clusters undergo expansion due to the relaxation processes.
The rate of this expansion notably depends on the cluster's relaxation timescale.
For example, the GCs with initial properties $M_{\rm GC} = 5\times 10^4\msun$ and $R_{\rm hm} = 0.7\pc$ undergoes an expansion of $R_{\rm hm}$ from $0.7\pc$ to $1.3\pc$.
In contrast, the GCs with initial properties $M_{\rm GC} = 10^5\msun$ and $R_{\rm hm} = 0.7\pc$ expands from $R_{\rm hm} = 0.7\pc$ to only $R_{\rm hm} = 1.0$, reflecting the relaxation timescale determined by the GC mass.

The evolution of the simulations in Group B is illustrated in Figure \ref{fig: LagR_Group_B}.
All simulations are initialized with the same $R_{\rm 5\%}$ and $R_{\rm hm}$ at $t = 0$ except for the small discrepancies caused by the adiabatic insertion of IMBH before initialization (See \ref{sec: adiabatic growth}).
As the GC evolves, the rate of expansion of the cluster size varies across different $M_{\rm IMBH,i}$.
For a lower-mass IMBH ($M_{\rm IMBH,i} = 300\msun$), the half-mass radius increases from $R_{\rm hm} = 0.7 \pc$ to $R_{\rm hm} = 1.1\pc$.
For a more massive IMBH ($M_{\rm IMBH,i} = 5000\msun$), the half-mass radius grows significantly more, reaching $R_{\rm hm} = 2.1\pc$.
This trend highlights the crucial role of IMBH mass in driving the dynamical expansion of the host GC as discussed in previous studies \citep[e.g.,][]{Baumgardt2004ApJ...613.1133B,Freitag2006JPhCS..54..252F,Hong2015MNRAS.448..754H}.

The detailed evolution of the cluster core is depicted in Figure \ref{fig: Core Evolution}.
The two left panels display the evolution of the central density, $\rho_{\rm c}$, for the populations within $r < R_{5\%}$ from the cluster's center.
In all simulations, the central density decreases monotonically as the core expands over time.
The evolution of the central density is significantly affected by both the initial star cluster properties and the mass of the central IMBHs.
This trend is already expected from the evolution of $R_{5\%}$ (See Figure \ref{fig: LagR_Group_A} \& \ref{fig: LagR_Group_B}).

The right panel of Figure \ref{fig: Core Evolution} presents the evolution of the central velocity dispersion, $\sigma_{\rm c}$.
Unlike the central density, the overall evolution of $\sigma_{\rm c}$ exhibits a more gradual and steady decline over time.
In Group A, the evolution of $\sigma_{\rm c}$ closely follows the trend of central density decline, supporting the interpretation that the decrease in velocity dispersion is primarily driven by the expansion of the cluster core.
As in the central density, the velocity dispersion of the clusters with longer $t_{\rm rlx}$ experiences more gradual dynamical evolution. 
For example, the GC with a total mass $M_{\rm GC} = 5\times 10^4\msun$ and an initial half-mass radius of $R_{\rm hm} = 0.5\pc$ undergoes a more than $30\%$ reduction in central velocity dispersion, decreasing from $20\km\s^{-1}$ to $13\km\s^{-1}$.
In contrast, the GCs with the same mass but a larger initial half-mass radius of $R_{\rm hm} = 1\pc$ experiences a relatively smaller ($\sim 20\%$) decline, with $\sigma_{\rm c}$ decreasing from $14 \km\s^{-1}$ to $11\km\s^{-1}$.

In Group B, the presence of an IMBH introduces additional dynamical effects.
The strong gravitational potential of a massive IMBH significantly enhances the velocity dispersion of stars near it, leading to distinct differences in the initial values of $\sigma_{\rm c}$ depending on the IMBH mass.
For an IMBH with $M_{\rm IMBH} = 5000\msun$, the initial central velocity dispersion is $\sim 24\km\s^{-1}$ at $t = 0\Myr$, gradually decreasing to $18\km\s^{-1}$ by $t = 30\Myr$ as cluster expands.
In contrast, for an IMBH with $M_{\rm IMBH,i} = 300\msun$, the initial velocity dispersion is significantly lower, starting at $\sim 14\km\s^{-1}$ at $t = 0\Myr$ and declining more modestly to $11 \km\s^{-1}$ by $t = 30\Myr$.
In summary, our result suggests that the evolution of the cluster center is highly governed by the properties of GCs, and the mass of the IMBH exerts a significant gravitational influence on the stars in its vicinity.

Figure \ref{fig: Mean_Mass} displays the evolution of the average mass of stars and SBHs within $R_{5\%}$ from the center.
The mean stellar mass exhibits a rapid increase early on and stabilizes by $10\Myr$ in all simulations.
In the top panel, the initial mean stellar mass differs between clusters of different total masses, being higher in the $5\times 10^4\msun$ GCs compared to the $10^5\msun$ cluster.
This discrepancy reflects a more rapid relaxation process during the adiabatic growth of IMBH prior to the initialization of simulations (See Section \ref{sec: adiabatic growth}).
Across all simulations in Group A, the central mean stellar mass stabilizes commonly at $\left<M_\star\right> \sim 0.9\msun$.
However, the subsequent evolution of the mean stellar mass depends on the initial properties of the star cluster, reaching values between $\sim 0.7\msun$ and $\sim 1\msun$ by $t = 30\Myr$.
Meanwhile, the lower panel highlights a clear decreasing trend in the average stellar mass at the cluster center with increasing IMBH mass.
This demonstrates a significant influence of IMBH mass on the mean stellar mass at the cluster center.

We now explore the implications of the initial conditions of a star cluster on the motion of an IMBH.
One key parameter is the IMBH's distance from the center, or the wandering radius, representing the typical distance the IMBH moves from the cluster center due to gravitational interactions with surrounding stars.
The wandering radius of the IMBH can be estimated using the following equation \citep{Stone2017MNRAS.467.4180S}:
\begin{eqnarray}
    d_{\rm IMBH} \approx R_{c} \left(\frac{\left<M_\star\right>}{M_{\rm IMBH}}\right)^{1/2}.
    \label{eq: d_IMBH}
\end{eqnarray}
While the equation originally uses $R_{\rm c}$ as the core size of the star cluster, there is ambiguity in how the core is defined for a star cluster in our simulations. 
Therefore, we instead adopt $R_{5\%}$ as a practical proxy.
The top-left panel of Figure \ref{fig: IMBH_Dynamics} confirms that the wandering radius increases with the size of the star clusters, consistent with our expectation that larger cores provide more room for the IMBH to move around.
The bottom-left panel, however, does not show a clear dependence of $d_{\rm IMBH}$ on $M_{\rm IMBH}$, even though Equation \ref{eq: d_IMBH} predicts an inverse-square-root scaling.

The principle of energy equipartition provides an estimate for the speed of the IMBH, $|\mathbf{v}_{\rm IMBH}|$ as follows:
\begin{equation}\label{eq: v_IMBH}
    |\mathbf{v}_{\rm IMBH}| = \sigma_{\rm c, 3D} \left(\frac{\left<M_\star\right>}{M_{\rm IMBH}}\right)^{1/2}
\end{equation}
where $\left<M_\star\right>$ is the average stellar mass, and $\sigma_{\rm c, 3D}$ is the 3D velocity dispersion of the stellar system given by $\sigma_{\rm c, 3D} = \sqrt{3} \sigma_{\rm c}$.
The two left panels in Figure \ref{fig: IMBH_Dynamics} illustrate the evolution of the velocity dispersion in Group A and Group B, respectively.
In Group A, $|\mathbf{v}_{\rm IMBH}|$ exhibits a mild decreasing trend with increasing GC size for fixed $M_{\rm GC}$.
Meanwhile, Group B shows a clear inverse dependence on $M_{\rm IMBH}$, consistent as expected by Equation \ref{eq: v_IMBH}.
However, the scales of IMBH's wandering radius and the measured speeds in both groups are found to be an order of magnitude larger than the estimates provided by Equations \ref{eq: d_IMBH} \& \ref{eq: v_IMBH}.
For example, inserting $M_{\rm IMBH,i} = 1000\msun$, $\left<M_\star\right> = 1.0\msun$, and $\sigma_{\rm c} = 14\km\s^{-1}$ (as seen for $M_{\rm IMBH,i} = 1000\msun$ in Group B) into Equation \ref{eq: v_IMBH}, we obtain $|v_{\rm IMBH}| \approx 1 \km\s^{-1}$.
This theoretical estimate is slightly lower than the measured value of $|v_{\rm IMBH}| \approx 3\km\s^{-1}$ presented in the bottom right panel of Figure \ref{fig: IMBH_Dynamics}.
This suggests that additional effects such as close encounters with SBHs might be influencing the IMBH's motion beyond the simple equipartition model.

The evolution of the GC cores in the real universe might not be as rapid as seen in these simulations, because the real GCs have longer relaxation timescales due to their greater masses.
The shortened relaxation time is an artifact of our simulation design as discussed in Section \ref{sec: sc parameters}.  
However, as we focus only on the central region surrounding the embedded IMBHs, and the statistical properties of IMRIs during the short runtime, we expect that our chosen GC properties would not significantly affect the conclusion from our analyses.

\vspace{1mm}

\subsection{The Statistical Properties of IMRI Events}\label{sec: IMRI Statistics}

In this section, we explore the rates and statistical properties of the IMRIs derived from our suite of $N$-body simulations.
Figure \ref{fig: IMRI Rate} depicts the IMRI event rates, $\Gamma_{\rm IMRI}$, in the GCs with initial 1D half-mass velocity dispersions ranging $\sigma_{\rm hm} \sim 6\,\mathrm{km}\,\mathrm{s}^{-1}$ to $\sim 10\,\mathrm{km}\,\mathrm{s}^{-1}$ (Group B in Table \ref{tab: Initial Conditions}).
They are calculated from the stellar and SBH populations within the half-mass radius in each simulation.
Each data point represents a value averaged over 10 different random initializations.
The initial mass of the IMBH is $M_{\rm IMBH,i}=1000\msun$.
Here we observe that $\Gamma_{\rm IMRI}$ increases with $\sigma_{\rm hm}$.  
The IMRI rate is approximately $\sim 0.06 \Myr^{-1}$ in a cluster system with $\sigma_{\rm hm} \sim 6\,\mathrm{km}\,\mathrm{s}^{-1}$, but it rises to $\sim 0.2\Myr^{-1}$ in a cluster with $\sigma_{\rm hm} \sim 10\,\mathrm{km}\,\mathrm{s}^{-1}$.
This trend suggests that denser clusters, characterized by higher velocity dispersions, are more conducive to the IMRI formation.

Figure \ref{fig: IMRI_Rate_MBH} highlights the relation between $M_{\rm IMBH}$ and $\Gamma_{\rm IMRI}$ for fixed cluster properties.
The IMRI event rate increases from $\sim 0.08\Myr^{-1}$ for a IMBH with $M_{\rm IMBH,i} = 300\msun$ to $\sim 0.2\Myr^{-1}$ for $M_{{\rm IMBH}, i} = 5000\msun$.
By integrating all our simulation results and fitting them to Eq.(\ref{eq: Linear_Fit_IMRI}), we obtain the best linear fit that estimates the IMRI event rate:
\begin{equation}
    \Gamma_\mathrm{IMRI} = 0.06\Myr^{-1}
    \left(\frac{\sigma_{\rm hm}}{6\,\mathrm{km}\,\mathrm{s}^{-1}}\right)^{2.72} \left(\frac{M_{{\rm IMBH},i}}{10^3\msun}\right)^{0.15}.
    \label{linear_fit}
\end{equation}
Note that this relation is built by employing a limited IMBH mass range from $300\msun$ to $5000\msun$.
Although it remains uncertain whether this relation can be extrapolated to the IMBH masses beyond this range, our linear-fit is sufficient to predict the lower mass IMBHs that would occupy the majority of the IMBH population, which is more sensitive to the nature of natal kick. See Section \ref{sec: detection rate discussion} for more information.

We now study the orbital properties of the IMRI events from our suite of $N$-body simulations.
The left panel of Figure \ref{fig: Encounters} displays a histogram of the orbital eccentricities found in all of our simulations of Group B (see Table \ref{tab: Initial Conditions}).
It shows that the majority of the IMRIs have high eccentricities, $\left(1 - e\right) \sim 10^{-3} - 10^{-4}$, at $1\Myr$ before the final merger.
Binaries with even higher eccentricities, $\left(1 - e\right) \lesssim 10^{-5}$, are less common because it is expected that such high values are more difficult to achieve in the first place.
Similarly, encounters with lower eccentricities, $\left(1 - e\right) \gtrsim 10^{-2}$, are uncommon because they need to achieve an extremely tiny semi-major axis to satisfy our merger criteria (i.e., only the binaries with $t_{\rm GW} < 1.0\,{\rm Myr}$ are merged in the simulation;  see Section \ref{sec: prescription}).
To validate this, in the middle panel of Figure \ref{fig: Encounters}  we show a relationship between the semi-major axes $a$ and the eccentricities $e$ from all Group B simulations.
Requiring $t_{\rm GW} = 1.0\Myr$ for a fixed combination of $M_{\rm IMBH}$ and $M_{\rm BH}$ provide us with a relation $a^4 \left(1-e\right)^{3.5} = \mathrm{constant}$ (see Eq.(\ref{eq: T_gw}) in  Section \ref{sec: prescription}; here we assume $\left(1 - e\right) \ll 1$).
Indeed, the black dashed line marks such a relation for $M_{\rm IMBH} = 10^3\msun$ and $M_{\rm BH} = 60\msun$ with  $t_{\rm GW}=1.0\Myr$. 
As expected, we observe that most data points are located below this line.
Lastly, the right panel of Figure \ref{fig: Encounters} presents the mass distribution of the secondary BHs involved in the IMRI event.
We find that an IMBH tends to form a binary with the most massive BHs in the star cluster.
%Given the rarity of such a massive BH population in the clusters, 

\vspace{1mm}

\section{Discussion} \label{sec:Discussion}

Here we estimate the actual detection rate of the IMRI events by the ground- and space-based GW observatories.

\begin{figure*}
    \centering
    \includegraphics[width=1.02\textwidth]{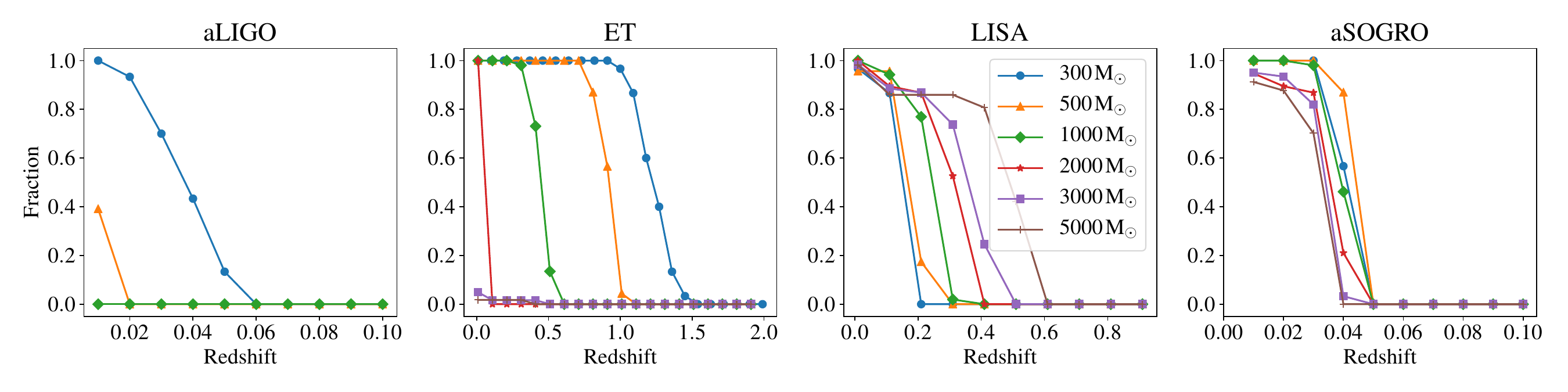}
    \vspace{-6mm}
    \caption{The fraction of IMRIs detectable ($S/N > 8$) by four observational instruments (aLIGO, ET, and LISA, and aSOGRO, from {\it left} to {\it right}) as a function of redshift, for IMBH masses, $M_{\rm IMBH}$, from 300 to $5000\msun$.
    The detectable redshift range varies significantly with the type of GW telescopes used. 
    aLIGO is capable of detecting very close GW sources at $z \lesssim 0.06$ only for $M_{\rm IMBH}=300\msun$. 
    In contrast, ET can potentially detect some IMRIs up to $z \sim 1.5$.
    Note that the maximum redshift of detection (``horizon redshift'') sharply decreases as $M_{\rm IMBH}$ increases. 
    LISA can detect IMRIs up to $z \sim 0.5$.  
    Unlike ET, the maximum redshift of detection increases as $M_{\rm IMBH}$ increases.
    aSOGRO can detect the wide mass range of the IMRIs up to $z \sim 0.04$.
    The plot for DECIGO is omitted, as it can detect all IMRI events across the entire range of redshifts and masses considered in this study.
    See Section \ref{sec: horizon discussion} for more information.}
    \label{fig: Detectability}
    \vspace{3mm}
\end{figure*}

\begin{figure}
    \centering
    \vspace{-3mm}      
    \includegraphics[width=0.93\columnwidth]{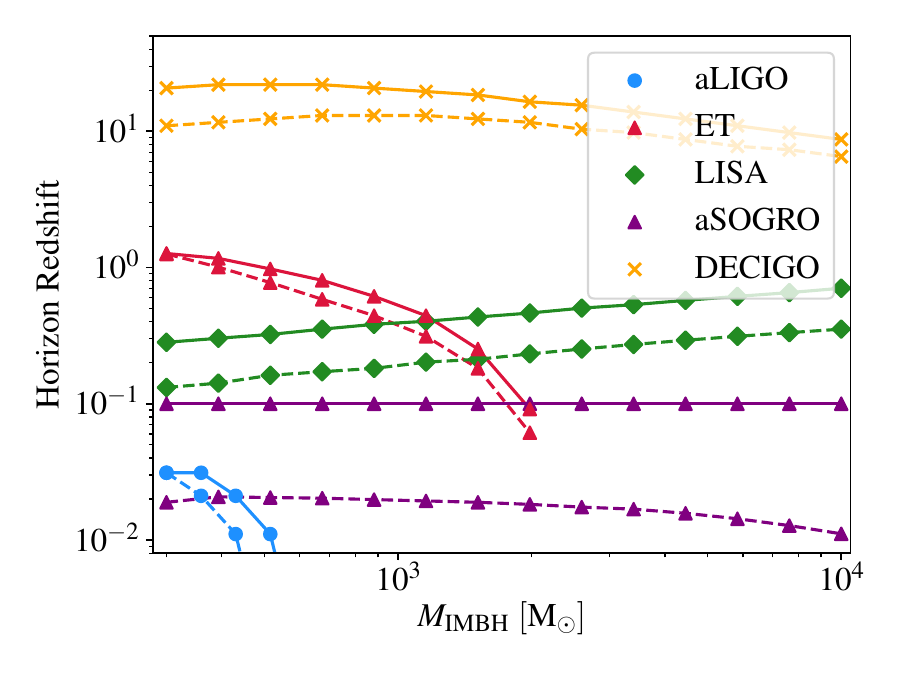}
    \vspace{-4mm}    
    \caption{The horizon redshift as a function of the IMBH mass for each GW observatory. 
    We assume that the secondary in the binary is a SBH of mass $10\msun$ ({\it dashed line}) and $60\msun$ ({\it solid line}), and that the eccentricities range from $e=0.999$ to $0.99999$ at $1\Myr$ before the merger.  
    The figure shows that aLIGO can only detect IMRIs involving $M_{\rm IMBH}  \lesssim 500\msun$ at $z \lesssim 0.06$. 
    ET has a significantly better capability for detecting these lower-mass IMBHs than aLIGO. 
    However, there is a dramatic kink at $M_{\rm IMBH} = 2000\msun$, meaning that ET is effective at detecting IMRIs involving IMBHs only with masses $M_{\rm IMBH} \lesssim 2000$. 
    Conversely, LISA and aSOGRO are capable of detecting IMBHs across all mass ranges in our analysis.
    However, LISA's horizon redshift grows from $z \sim 0.2$ to $z \sim 0.5$ with $M_{\rm IMBH}$ increased, while aSOGRO's is limited within $z \lesssim 0.05$.
    DECIGO possesses the widest detection range among the five observatories in our study.
    Ideally, it would detect all IMRIs from the entire universe. 
    See Section \ref{sec: horizon discussion} for more information.}
    \label{fig: Horizon Redshift}
    \vspace{1mm}    
\end{figure}

\subsection{Horizon Redshifts for IMRI Detection} \label{sec: horizon discussion}

Figure \ref{fig: Detectability} presents the fraction of observable IMRI events for different GW telescopes, assuming events are detectable if its SNR is larger than 8 (Eq.(\ref{eq: S/N}) in Section \ref{sec: snr}).
In the first panel, aLIGO is unable to detect IMBHs with $M_{\rm IMBH} \gtrsim 500\msun$ because its detection band is limited to frequencies $\gtrsim 10\,\mathrm{Hz}$.
Even for $M_{\rm IMBH} \lesssim 300\msun$, it can only detect very close IMRIs in the local universe ($z \lesssim 0.06$).
Consequently, we can reasonably conclude that aLIGO would not be able to discover a merger event involving an IMBH, unless there is one with mass $\lesssim 300\msun$ in the vicinity of Milky Way ($\lesssim 260 \Mpc$).  

ET has a better capability to detect IMRIs compared to aLIGO, especially in the lower-mass range of $M_{\rm IMBH}$.  
For example, the second panel of Figure \ref{fig: Detectability} shows that it has the potential to detect almost all $300\msun$ IMBHs up to $z \sim 1.0$, and some up to $z \sim 1.5$.
ET also offers a relatively wider detection mass spectrum up to $M_{\rm IMBH} \sim 2000\msun$, although the horizon redshift decreases to $z \lesssim 0.1$.
Note that the maximum redshift of detection --- or the ``horizon redshift'' --- decreases as $M_{\rm IMBH}$ increases.

On the other hand, we find that LISA, aSOGRO, and DECIGO can observe all IMRIs involving IMBHs across the entire mass range explored.
However, their horizon redshifts vary significantly from each other.
Figure \ref{fig: Horizon Redshift} illustrates the horizon redshifts of each GW observatory across the IMBH masses $M_{\rm IMBH} = 300\msun$ to $10^4\msun$.
LISA has a growing horizon redshift for increasing $M_{\rm IMBH}$ from $z \sim 0.2$ to $\sim 0.5$.
Especially, it can better cover the detection of IMRIs including $M_{\rm IMBH} \gtrsim 2000\msun$ than ET.
Meanwhile, aSOGRO's horizon redshift is confined within the local universe $z \lesssim 0.05$.
Although its detection range diminishes for higher mass IMBHs, DECIGO can detect IMRI events occurring at redshifts as high as $z \sim 10$.

In conclusion, both aLIGO and DECIGO may serve as a local probe for low-mass IMBHs, but aSOGRO covers a broader mass range.
Meanwhile, ET and LISA can extend the cosmic volume under our detection up to $z \sim 0.5 - 1.0$, complementing each other in their detection capabilities across different mass ranges. 
Lastly, DECIGO can push the observational limits beyond $z \sim 10$ across all mass ranges.

\subsection{IMRI Detection Rates By GW Observatories} \label{sec: detection rate discussion}

We now estimate the IMRI detection rates by three different GW observatories, employing the GC number density as discussed in Section \ref{sec: IMRI Detection Rate}.
In this work, the distribution of half-mass velocity dispersions, $\sigma_{\rm hm}$, of GCs is derived from
\begin{equation}
    \frac{\mathrm{d} \nu_{\rm GC}}{\mathrm{d} \sigma_{\rm hm}} = \frac{\mathrm{d} \nu_{\rm GC}}{\mathrm{d} M_{\rm GC}}\frac{\mathrm{d} M_{\rm GC}}{\mathrm{d} \sigma_{\rm hm}}.
    \label{eq: dndsigma}
\end{equation}
Here, we use the relation $\sigma_{\rm hm} = \sqrt{G M_{\rm GC}/6 R_{\rm hm}}$ to estimate velocity dispersion, assuming a typical GC size of $R_{\rm hm} \sim 2.5\pc$, consistent with observations of GCs in the Milky Way \citep{Portegies2010ARA&A..48..431P, Shin2013ApJ...762..135S}.

\begin{figure}
    \centering
    \includegraphics[width=0.91\columnwidth]{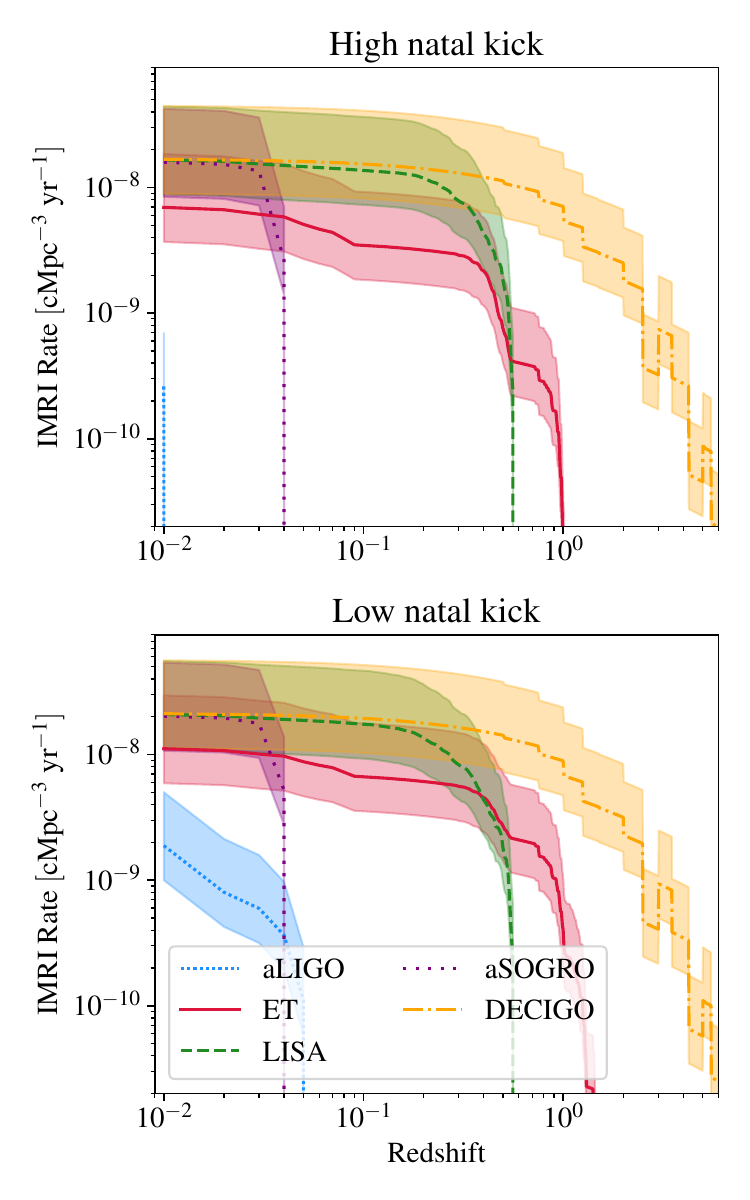}
    \vspace{-3mm}
    \caption{The prospect of IMRI detection rates per unit comoving volume at each redshift for aLIGO ({\it densely-dotted blue line}), ET ({\it solid red line}), LISA ({\it dashed green line}), aSOGRO ({\it loosely-dotted purple line}), and DECIGO ({\it dash-dotted yellow line}) under scenarios of high ({\it top}) and low ({\it bottom}) natal kicks (Section \ref{sec: IMRI Detection Rate}). 
    The shaded region represents the range of detection rates corresponding to varying GC number densities, $n_{\rm GC} = 0.77 - 3.77\Mpc^{-3}$ (Section \ref{sec: IMRI Detection Rate}).
    See Section \ref{sec: detection rate discussion} for more information.}        
    \vspace{1mm}    
    \label{fig: Conv_Rate}
\end{figure}

Combining and integrating Eqs.(\ref{eq: IMBH-GC relation}), (\ref{linear_fit}) and (\ref{eq: dndsigma}) yields the expected IMRI detection rates per unit comoving volume at each redshift.
Figure \ref{fig: Conv_Rate} shows the resulting detection rates for two natal kick scenarios (see Section \ref{sec: IMRI Detection Rate}).
In both scenarios, all observatories except for aLIGO have the rate $\sim$ a few $\times 10^{-8}\,\mathrm{cMpc}^{-3}\yr^{-1}$ at $z \lesssim 0.3$.
First of all, its detection range encompasses the entire history of IMRI mergers, since DECIGO may cover all events up to a redshift $\lesssim 10$.
LISA can detect some IMRIs up to the $z \lesssim 0.5$, while ET's detection range extends slightly farther, up to $z \lesssim 1.0$.
Despite their similar detection rates at $z < 0.3$, it is important to note that they are complementary in terms of the mass spectrum they can observe (see Figure \ref{fig: Horizon Redshift} and Section \ref{sec: horizon discussion}).

Meanwhile, aLIGO and aSOGRO have much narrower detection ranges ($z \lesssim 0.06$); However, aLIGO exhibits a remarkably lower detection rate per unit comoving volume of $\lesssim 2 \times 10^{-9} \, \mathrm{cMpc}^{-3}\yr^{-1}$ compared to aSOGRO.
This gap can be attributed to aLIGO's limited detectable mass range even in the optimistic scenario with low natal kicks.
In the pessimistic-case scenario in which many lower-mass IMBHs are ejected due to strong natal kicks, aLIGO will barely detect any IMRIs, even in very close proximity ($z < 0.01$).
Figure \ref{fig: Conv_Rate} illustrates that, for some GW telescopes, the IMRI detection rate can be susceptible to the intensity of natal kicks assumed in our model. 
    
By integrating the above results up to the horizon redshift, we derive the final conclusions of our study --- that is the IMRI detection rates per year for different GW telescopes under two scenarios of BH natal kicks.
The detection rate for aLIGO is found as
\begin{equation}
    \Gamma_{\rm det, \,aLIGO} \sim 1.1\times 10^{-4} - 0.02 \yr^{-1} \left(\frac{n_{{\rm GC}, \,z = 0}}{1.4\Mpc^{-3}}\right),
\end{equation}
where we include the dependency on $n_{{\rm GC}, \,z = 0}$ in the formula.
The aLIGO mainly targets low-mass IMBHs which are vulnerable to the recoil kick in general \citep{Heggie1975MNRAS.173..729H, Antonini2016ApJ...831..187A, Arca-Sedda2023MNRAS.526..429A}.
This makes it challenging to constrain the detection frequencies for IMBHs with $M_{\rm IMBH} < 10^3\msun$.
In extreme cases, these IMBHs may be completely ejected from their host GCs during evolution.
This uncertainty is reflected in Table 1 of \citet{Askar2022MNRAS.511.2631A}, which shows the highest variance in the retention of low-mass IMBHs.
In summary, our results indicate that aLIGO will likely detect very few IMBHs, even under the most optimistic assumptions.
    
Meanwhile, the detection rate of upcoming GW observatories will significantly improve the chance of IMRI detection.
The detection rate of the rest of GW observatories are as follows. 
The rate of ET is estimated as
\begin{equation}
    \Gamma_{\rm det, \, ET} \sim 101 - 355\yr^{-1} \left(\frac{n_{{\rm GC}, \,z = 0}}{1.4\Mpc^{-3}}\right),
\end{equation}
offering ample chances of IMRI detection for $M_{\rm IMBH} \lesssim 10^3 \msun$.
The detection rate of LISA is
\begin{equation}
    \Gamma_{\rm det, \, LISA} \sim 186 - 200 \yr^{-1} \left(\frac{n_{{\rm GC}, \,z = 0}}{1.4\Mpc^{-3}}\right).
\end{equation}
LISA appears to be less affected by the natal kicks, as more massive IMBHs dominate its detectable IMRI populations (see Figure \ref{fig: Horizon Redshift} and Section \ref{sec: horizon discussion}).
The detection rate of aSOGRO is given as
\begin{equation}
    \Gamma_{\rm det, \, aSOGRO} \sim 0.24 - 0.34 \yr^{-1} \left(\frac{n_{{\rm GC}, \,z = 0}}{1.4\Mpc^{-3}}\right).
\end{equation}
Lastly, the detection rate of DECIGO is estimated as
\begin{equation}
    \Gamma_{\rm det, \, DECIGO} \sim 3880 - 4890 \yr^{-1} \left(\frac{n_{{\rm GC}, \,z = 0}}{1.4\Mpc^{-3}}\right)
\end{equation}
making it the most promising GW observatory for detecting IMRIs among those considered in this study.

According to our analyses, ET, LISA, and DECIGO are expected to provide robust detections of IMRIs, thereby greatly improving our ability to study the IMBH population over a wide range of masses and redshifts. 
Although aSOGRO exhibits a lower detection rate compared to other future observatories, it remains capable of detecting IMRI events within a practical observational timescale ($\sim 5 - 10 \yr$).

\subsection{Caveats In Our Analyses} \label{sec: Caveats}

In this section, we discuss the limitations of our study, focusing on the factors that are not considered in our analyses.
\begin{itemize}
    \item Our simulations assume that the IMBH-BH binary tightens only via GW radiation.
    However, a binary may undergo multiple encounters with other stars or SBHs nearby; and, these ``third'' objects passing close to the center of mass of the hard binary can deform the binary configuration \citep{Saslaw1974ApJ...190..253S, Merritt2005LRR.....8....8M}.
    As a result, on average, the actual merger timescale could be shorter than our simulation results.
    More importantly, these interactions can significantly alter the characteristics of the observed GW signals.
    Also, the general relativistic effect such as the capture by gravitational capture \citep{Quinlan1989ApJ...343..725Q} is not included in our simulation.
    Given the rare frequency of the gravitational capture events, this would not affect our result significantly (See Appendix \ref{Appendix: GR capture timescale}).

    \item Our study also primarily focuses on the GW signals during the inspiral stage since it is the dominant source.
    However, despite its short duration, the merger and ringdown phases can also produce loud signals in some cases.
    Consequently, aLIGO and ET may have the potential to observe a broader mass spectrum than our predictions.

    \item In a realistic GW observation, multiple aspects can affect such as the Earth rotation \citep[e.g.,][]{Isi2015PhRvD..91h2002I}. 
    However, the SNR calculation utilizing Eq.(\ref{eq: S/N}) will not be significantly different from the precise measurement as an order of magnitude sense.
    Therefore, the actual IMRI detection rate will not be greatly far from our estimation.
    
    \item The time evolution of IMRI rates may depend on the formation mechanism of the IMBH. 
    IMBHs can form through two distinct scenarios: the slow and fast scenarios \citep{Giersz2015MNRAS.454.3150G}.
    In the slow scenario, an IMBH grows gradually over time by accreting stars and BHs, allowing IMRI events persists up to the present day.
    In contrast, in the fast scenario, where the IMBH forms via the runaway merger of massive stars, the majority of IMRI events are expected to occur within the first $5\Gyr$, depleting the sources of IMRIs by low redshifts \citep{Hong2020MNRAS.498.4287H}.

    \item The mass range of IMBHs is constrained to $M_{\rm IMBH} < 10^4 \msun$ in our analysis.
    However, observatories such as LISA and DECIGO may be also sensitive to $M_{\rm IMBH} > 10^4\msun$.
    Consequently, the actual detection rates for these two observatories are likely to exceed the prediction of our analysis.

    \item The low-mass IMBHs are vulnerable to the GW recoil kicks from the IMBH-BH merger \citep{Gonzalez2022ApJ...940..131G}.
    The GW kick may eject most of the low-mass IMBH from the dense star clusters, and this would result in the decrease of the IMRI rate for IMRIs with $\left(M_{\rm IMBH} < 10^3\msun \right)$.
    The impact of GW kick on the IMRI detection rate will be rather limited since detectors such as LISA and aSOGRO are sensitive to higher mass regimes.
    Meanwhile, ET and aLIGO may detect less frequently than our estimations although we already concluded that aLIGO would be less likely to detect any IMRIs in a practical timescale.

    \item The detailed evolution of the IMRI rate across cosmic time may exhibit significant variations depending on the assumptions on the GC formation and evolution history.
    For example, a semi-analytic model work of \citet{Fragione2018ApJ...867..119F} accounts for the mass loss of GCs due to the galactic tidal field. 
    Their study predicts that at low redshift, the IMRI event rate falls within the range of $\sim 0.5 - 3 \Gpc^{-3}\yr^{-1}$. 
    However, during the peak GC formation epoch, this rate increases as high as $\sim 100\Gpc^{-3}\yr^{-1}$.
    In contrast, our simulations, which focus on short-term GC evolution, predict an order of magnitude higher IMRI rates than those found in \citep{Fragione2018ApJ...867..119F}, while it declines at higher redshifts because of the decreasing GC number density.
\end{itemize}

\section{Summary and Conclusion}\label{sub: Summary}

Using a suite of direct $N$-body simulations we have studied the IMRI event rates, while varying the velocity dispersion of the embedding GCs and the mass of the central IMBHs.
We have also explored the distribution of the orbital parameters of the simulated IMBH-BH binaries.
Our simulation study thus provides a comprehensive overview of the IMRI statistics, namely, the event rates and the orbital properties (Section \ref{sec:Results}).
With this information, we have estimated the detection rates of IMRI events involving IMBHs by the current and future ground-/space-based GW observatories (Section \ref{sec:Discussion}).
Our main results are summarized as follows:
\begin{itemize}
    \item The IMRI event rate typically falls in the range of  $0.1 - 0.3\,\Myr^{-1}$ for the IMBHs of mass $M_{\rm IMBH}=300 - 5000\msun$. 
    The IMRI event rate is influenced by both the velocity dispersion of the embedding GCs and the mass of the IMBH (Section \ref{sec: IMRI Statistics}).
    
    \item LISA can detect IMRIs involving IMBHs across a wide range of IMBH masses, with its horizon redshift increasing as $M_{\rm IMBH}$ increases.
    It can detect the majority of eccentric IMRIs in the local universe, and even some IMRIs with $M_{\rm IMBH}= 5000\msun$ up to $z \sim 0.5$.
    Although ET's detection capability is restricted to IMRIs with  $M_{\rm IMBH} \lesssim 2000\msun$, it serves as a complementary instrument to LISA within this mass spectrum.
    DECIGO extends GW observational range beyond $z\sim 10$, covering the entire mass ranges considered in our analysis.
    In contrast, aSOGRO's horizon redshift is limited within the Milky Way's vicinity as aLIGO's (Section \ref{sec: horizon discussion}).
    Nevertheless, aSOGRO is capable of covering a broader mass spectrum than aLIGO.
    
    \item Integrating the above results in the redshift space, we derive the IMRI detection rate per year for three GW observatories: $\lesssim  0.04\yr^{-1}$ for aLIGO, $\sim 101 - 355 \yr^{-1}$ for ET, $\sim 186 - 200 \yr^{-1}$ for LISA, $\sim 0.24 - 0.34 \yr^{-1}$ for aSOGRO, and $\sim 3880 - 4890 \yr^{-1}$ for DECIGO.
    This result can be affected by the (local) GC number density or the IMBH retention fraction (Section \ref{sec: detection rate discussion}).
\end{itemize}

Our study provides a pipeline to investigate the IMRI detection rates for GW observatories, by utilizing direct $N$-body simulations on dense stellar systems.
Our results demonstrates that the upcoming GW observatories will have several chance of observing the IMRIs.
With the improved observational constraints in the future, such as the local GC density and BH natal kicks, our pipeline will provide a comprehensive perspective on the robust detection of IMBHs.

Other than the GW radiated from IMRIs explored in this study, there are several observable signals for IMBHs.
One can be the optical flare from the tidal disruption of stars or white dwarfs by IMBHs \citep{Strubbe2009MNRAS.400.2070S}.
If detected, the IMBH-IMBH merger can be considerably manifest \citep{Rasskazov2020ApJ...899..149R}.
In addition, a hypervelocity star escaping from a star cluster can be another indirect evidence of the IMBH \citep{Fragione2019MNRAS.489.4543F}.
We leave the study of providing a more realistic IMBH detection rate by considering all of these factors as our future project.

\begin{acknowledgments}
Seungjae Lee would like to thank Chan Park, and Boon Kiat Oh for their kind and enlightening comments on the early version of this manuscript.
We are grateful to Han Gil Choi who kindly introduced us a detailed background information on the GW detections.
We also thank Sambaran Banerjee, Francesco Maria Flammini Dotti and Kai Wu for their insightful discussions and detailed comments on the {\it N}-body simulations.
Hyumg Mok Lee was supported by National Research Foundation of Korea (NRF) grant funded by the Korea government (MSIT) (No. 2021M3F7A1082056).
Ji-hoon Kim’s work was supported by the NRF grant (No. 2022M3K3A1093827 and No. 2023R1A2C1003244). 
His work was also supported by the Global-LAMP Program of the NRF grant funded by the Ministry of Education (No. RS-2023-00301976).
His work was also supported by the National Institute of Supercomputing and Network/Korea Institute of Science and Technology Information with supercomputing resources including technical support, grants KSC-2021-CRE-0442, KSC-2022-CRE-0355 and KSC-2024-CRE-0232.
\end{acknowledgments}

\appendix
\section{Hardness of the binary}\label{Appendix: hardness of binary}
The hardness of an IMBH-BH binary can be quantified as $H = \left| E_b \right|/E_k$, where $E_b= -G M_{\rm IMBH} M_{\rm BH}/\left(2a\right)$ represents the binding energy of the binary, and $E_k = M_{\rm third} v_{\rm third}^2/2$ is the typical kinetic energy of surrounding stars and BHs.
Here, $M_{\rm third}$ is the mass of an incoming third particle, and $v$ is its velocity.
Using the parameters $M_{\rm IMBH} = 1000 \msun$, $M_{\rm BH} = 20 \msun$, $a = 10 \au$, $M_{\rm third} = 10\msun$, $v_{\rm third} = 10\km \s^{-1}$ gives us $H \approx 1800$. 
Refer to Table \ref{tab: Initial Conditions} and Figure \ref{fig: Encounters} for the choice of those parameters.
This high hardness value indicates that the binary is sufficiently resistant to disruption from third-body encounters.

\section{The multi-body interaction timescale}\label{Appendix:  encounter timescale}
The timescale of encounters between an IMBH-BH binary and a third body can be estimated by 
\begin{equation}
    t_{\rm encounter} = 1/(n \Sigma_{\rm b} v)
    \label{eq: t_enc}
\end{equation}
where $n$ is the number density of stars and SBHs in the region surrounding the binary, $\Sigma_{\rm b}$ is the cross-section of the binary, and $v$ is the speed of the encountering object, or the velocity dispersion. 
Inserting $n = 10^5 \pc^{-3}$, $\Sigma_{\rm b} = \pi R_{\rm b}^2$, and $v = 7\km\,\s^{-1}$ gives us $t_{\rm encounter} \approx 45\Myr$ (See Table \ref{tab: Initial Conditions}).
Here, we choose the radius of the binary to be $R_{\rm b} = 10^{-4} \pc$ considering the merger criteria in our simulations.
This result indicates that an IMBH-BH binary is unlikely to encounter a third-body within the GW merger timescale of $1 \Myr$ assumed in our model.

\section{The gravitational capture timescale}\label{Appendix: GR capture timescale}
When two black holes become close, gravitational radiation may dissipate most of the orbital kinetic energy, thus forming a binary.
The timescale of the gravitational capture can be estimated in the similar manner as Equation \ref{eq: t_enc}:
\begin{equation}
    t_{\rm DG} = 1/\left(n \Sigma_{\rm GR,cap} v\right)
    \label{eq: t_DG}
\end{equation}
where $\Sigma_{\rm DG}$ is the cross-section of the gravitational capture. 
Exploiting Equation (15) of \citet{Lee1995MNRAS.272..605L}, we estimate $\Sigma_{\rm DG}$ as follows:
\begin{equation}
    \Sigma_{\rm DG,cap} = 2\pi\left(\frac{85\pi}{6\sqrt{2}}\right)^{2/7} \frac{G^2 M_{\rm IMBH}^{2/7} M_{\rm BH}^{2/7}\left(M_{\rm IMBH} + M_{\rm BH}\right)^{10/7}}{c^{10/7} v^{18/7}}.
\end{equation}
Substituting $n = 10^5\pc^{-3}$, $M_{\rm IMBH} = 10^3\msun$, $M_{\rm BH} = 10\msun$, and $v = 15\km\s^{-1}$ into Equation \ref{eq: t_DG} and \ref{eq: b_DG} yields a direct merger timescale of $t_{\rm DG} \approx 550 \Myr$, corresponding to a merger rate of $\lesssim 0.002\Myr^{-1}$.
This result demonstrates that the direct gravitational capture event would be rarer than IMRIs driven by gravitational wave dissipation.

\section{Details of the {\it N}-body Units}\label{Appendix: unit}
All the {\it N}-body simulations use dimensionless units \citep{Aarseth2003gnbs.book.....A}.
For a star cluster with initial total particle number $N$, and the mean particle mass, $\bar{M}$ in $\msun$, and equilibrium virial radius, $\bar{R}$ in parsec, are set to be unity with the total energy $E_0 = -0.25$ for bound systems.
Then the physical unit of velocity and time can be expressed as
\begin{align}
    \bar{V} &= 6.557\times 10^{-2}\left(\frac{N\, \bar{M}}{\bar{R}}\right) \km \s^{-1},\\
    \bar{T} & = 14.94\left(\frac{\bar{R}^3}{N\, \bar{M}}\right)^{1/2} \Myr.
\end{align}
The viral radius of the Plummer model, which is adopted in our experiments, is approximately estimated by $\bar{R} = 1.3 R_{\rm hm}$

\section{The Schechter Function} \label{Appendix: schechter}

The number density of the galaxies can be obtained by integrating the \citet{Schechter1976ApJ...203..297S} function.
The Schechter function for the number densities of galaxies is given by
\begin{equation}
    \phi(\Bar{M}) = \phi^* \ln(10) \left[10^{b(\Bar{M}-M^*)}\right]^{(1 + \alpha)}\exp{\left(-10^{b(\Bar{M}-M^*)}\right)},
\end{equation}
where the parameters $\phi^*$, $\alpha$, $M_\star$, and $M^*$ are from  Table 1 of \cite{Conselice2016ApJ...830...83C} and redshift-dependent, and $\Bar{M} = \log{M_\star}$.
The total number densities of the galaxies at a given redshift can be estimated by integrating this function as
\begin{equation}
    \phi_{\rm T}\left(z\right) = \int^{\Bar{M}_2}_{\Bar{M}_1}\phi(\Bar{M},z)d\Bar{M},
\end{equation}
whose integrated form can be well approximated by 
\begin{equation}
    \phi_{\rm T}\left(z\right) \approx \frac{-\phi^* 10^{(\alpha + 1)(M_2 - M^*)}}{\alpha + 1}.
\end{equation}
We also assume that the stellar masses of galaxies are all identical to $6\times 10^{10}\msun$. 
The total number of galaxies at $z < z_0$ can be given by the integral
\begin{equation}
    N_{\rm tot} = \int^{t_0}_{0}\int^{4\pi}_0 D_H \frac{\left( 1+z \right)^2 D_A^2}{E\left(z\right)}\phi_T (t) d\Omega dt,
\end{equation}
where $D_A$ is the angular size distance, $D_H = c/H_0$, and $E\left(z\right) = (\Omega_M\left(1+z\right)^3 + \Omega_\lambda )^{1/2}$.

\section{Evolution of the IMBH-BH binary} \label{Appendix: IMRI Evolution}

The orbit-averaged evolution of the semi-major axis $a$ and eccentricity $e$ is given by  \citep{Peters1964PhRv..136.1224P}: 
\begin{equation}
    \left<\frac{\mathrm{d}a}{\mathrm{d}t}\right> = -\frac{64}{5}\frac{G^3 M_1 M_2 \left(M_1 + M_2\right)}{c^5 a^3 \left(1-e^2\right)^{7/2}}\left(1 + \frac{73}{24}e^2 + \frac{37}{96}e^4\right)
\label{eq: dadt}
\end{equation}
and
\begin{equation}
    \left<\frac{\mathrm{d}e}{\mathrm{d}t}\right> = -\frac{304}{15} e\frac{G^3 M_1 M_2 \left(M_1 + M_2\right)}{c^5 a^4 \left(1 - e^2\right)^{5/2}}\left(1 + \frac{121}{304}e^2\right).
\label{eq: dedt}
\end{equation}
Additionally, the time-averaged energy emission rate is 
\begin{equation}
    \left<\frac{\mathrm{d} E}{\mathrm{d} t}\right> = -\frac{32}{5} \frac{G^4 M_1^2 M_2^2\left(M_1 + M_2\right)}{c^5 a^5\left(1 - e^2\right)^{7/2}}\left(1 + \frac{73}{24}e^2 + \frac{37}{96}e^4\right),
\end{equation}
while their merger timescale is estimated using Eq.(\ref{eq: T_gw}).

Following Eqs.(19) - (20) of \citet{Peters1963PhRv..131..435P}, the power radiated through the $n$-th harmonic can be computed as
\begin{equation}
    \dot{E}_n = \frac{32}{5} \frac{G^4 M_1^2 M_2^2\left(M_1 + M_2\right)}{c^5 a^5} g\left(n,e\right),
    \label{eq: E_n}
\end{equation}
where
\begin{eqnarray}
g\left(n,e\right) &=& \frac{n^4}{32} \Bigg\{ \bigg[ J_{n-2}\left(ne\right) - 2e J_{n-1}\left(ne\right)+\frac{2}{n}J_n\left(ne\right) \nonumber \\
&+&2eJ_{n+1}\left(ne\right) - J_{n+2}\left(ne\right)\bigg]^2 +\left(1-e^2\right)\big[J_{n-2}\left(ne\right) \nonumber \\
&-&2 J_n\left(ne\right) + J_{n+2}\left(ne\right)\big]^2+ \frac{4}{3n^2}\left[J_n\left(ne\right)\right]^2 \Bigg\},
\end{eqnarray}
and $J_n$ are Bessel functions of the first kind.

\section{Number Density of Globular Clusters} \label{Appendix: n_GC}

The total stellar mass contained in a GC can be approximated as
\begin{equation}
    M_{\rm tot, GC} = S_M {M_{\mathrm{G}_\star}}
\end{equation}
with the specific mass $S_M \approx 0.001 - 0.003$ and the total bulge stellar mass $M_{\rm G_\star}$ \citep{Peng2008ApJ...681..197P,Harris2013ApJ...772...82H}.
Then we can obtain the GC number density by the integral
\begin{equation}
     n_{\rm GC}(z) = \int^{\Bar{M}_2}_{\Bar{M}_1}\phi(\Bar{M},z)N_{\rm GCs}(\Bar{M}) d \Bar{M},
\end{equation}
where the number of GCs in a galaxy, $N_{\rm GCs}(\Bar{M}) = 10^{-4} M_\star/\left<M_{\rm GC}\right>$.

\section{The sensitivity curves of detectors}  \label{Appendix: sensitivity}
We model the sensitivity curves of the five detectors: aLIGO,\footnote{https://dcc.ligo.org/cgi-bin/DocDB/DocumentDatabase} ET,\footnote{\citet{Hild2011CQGra..28i4013H}} LISA, aSOGRO, and DECIGO.
The sensitivity of LISA is estimated using Eq.(13) of \citet{Robson2019CQGra..36j5011R}, as 
\begin{eqnarray}
    S_{h, \mathrm{LISA}}\left(f\right) = \frac{10}{3L_{\rm LISA}^2}\left(P_{\rm OMS}\left(f\right) + \frac{4 P_{\rm acc}\left(f\right)}{\left(2\pi f\right)^4} \right)
\nonumber
\\
    \times \left(1 + \frac{3}{5}\left(\frac{f}{f_\star}\right)^2\right) + S_{\rm c}\left(f\right),
\end{eqnarray}
where $L_{\rm LISA} = 2.5\times 10^{11}\cm$ is the arm length of LISA, $f_\star = 19.09\,\mathrm{mHz}$, $P_{\rm OMS}\left(f\right)$ is the optical metrology noise, $P_{\rm acc}\left(f\right)$ is the acceleration noise, and $S_{\rm c}\left(f\right)$ is the confusion noise.
For details of this expression, we refer the interested readers to \citet{Robson2019CQGra..36j5011R}.
In the present study, we ignore the contribution of the confusion noise.

The sensitivity of aSOGRO is drawn based on the Eq.(13) and Table.(1) of \citet{Bae2024PTEP.2024e3E01B}, as
\begin{eqnarray}
    S_{h, \mathrm{SOGRO}}(f) &=& \frac{16}{M_{\rm TM} L_{\rm aSOGRO}^2 \omega^4}\\
    &\times& \left[\frac{k_B T \omega_D}{Q_D} + \frac{|\omega^2 - \omega^2_D|}{\omega_p}\left(1 + \frac{1}{\beta^2}\right)^{1/2}k_B T_N\right].\nonumber 
\end{eqnarray}
with $\omega = 2\pi f$ is the angular frequency, $k_B$ is the Boltzmann constant, $M_{\rm TM} = 5000\,\mathrm{kg}$ is the mass of the individual freely moving test masses, and $L_{\rm aSOGRO} = 50\,\mathrm{m}$ is the arm length of aSOGRO, respectively;
$T = 0.1\,\mathrm{K}$ is the Antenna temperature;
$Q_D = 10^8$ and $\omega_D = 2 \pi \times 10^{-2}\,\mathrm{Hz}$ is the angular resonance frequency and quality factor of differential mode (DM);
$\omega_p = 2\pi \times 50\,\mathrm{kHz}$ is the pump (angular) frequency;
$T_N \equiv 5 \hbar \omega_P/k_B$ is the noise temperature.
The energy coupling constant $\beta$ is computed by
\begin{eqnarray}
    \beta = \frac{2C E_p^2 Q_p}{M_{\rm TM}|\omega^2 - \omega^2_D|}\frac{1}{\sqrt{1 + \left(2 Q_p \omega/ \omega_p\right)^2}}
\end{eqnarray}
where $C = 4\times 10^{-9}\,\mathrm{F}$ is the equilibrium value of each sensing capacitor, $E_p = 5\times 10^5\,\mathrm{V}\,\mathrm{m}^{-1}$ is the amplitude of the driving electric at $\omega_p$, and $Q_p = 10^6$ is platform quality factor.

The sensitivity curve of DECIGO is given by the Eq.(5) of \citep{Yagi2011PhRvD..83d4011Y}, as 
\begin{eqnarray}
    S_{h, \mathrm{DECIGO}}(f) &=& 6.53\times 10^{-49}\left[1 + \left(\frac{f}{7.36 \,\mathrm{Hz}}\right)^2\right] \nonumber \\
    &+& 4.45\times 10^{-51}\left(\frac{f}{1\,\mathrm{Hz}}\right)^{-4}\left[1 + \left(\frac{f}{7.36 \,\mathrm{Hz}}\right)^2\right]^{-1} \nonumber \\
    &+& 4.94\times 10^{-52} \left(\frac{f}{1\,\mathrm{Hz}}\right)^{-4}\,\mathrm{Hz}.
\end{eqnarray}

\bibliography{main}{}
\bibliographystyle{aasjournal}



\end{document}